\documentclass[11pt,a4paper]{article} 
\pdfoutput=1
\usepackage{jcappub} 
\usepackage{verbatim}
\usepackage{listings}
\RequirePackage[T1]{fontenc}
\usepackage{shadow}
\usepackage{varioref}
\usepackage{array}
\usepackage{rotating}
\usepackage{multirow}
\usepackage{hhline}

\usepackage{floatrow}
\usepackage[strict]{changepage}

\usepackage{myaasmacros}

\RequirePackage{graphicx}
\RequirePackage{mathptmx}      % use Times fonts if available on your TeX system
\RequirePackage{flushend}
\RequirePackage[numbers,sort&compress]{natbib}

\usepackage{amsmath}  
\usepackage{amssymb}  
\usepackage{subfig}  
\usepackage{graphicx}  
\usepackage{paralist}  
\usepackage{multirow}  
\usepackage{amsbsy}  
\usepackage{amsfonts}  
\usepackage{cuted}  
\usepackage{xspace}  
\PassOptionsToPackage{numbers,sort&compress}{natbib}  
\usepackage{lmodern}

\title{Generic energy transport solutions to the solar abundance problem -- a hint of new physics}

\author[a,b]{A.V. Sokolov}

\affiliation[a]{Institute for Nuclear Research of the Russian Academy of Sciences, 117312 Moscow, Russia}
\affiliation[b]{Deutsches Elektronen Synchrotron, Notkestrasse 85, 22607 Hamburg, Germany}

\emailAdd{anton.sokolov@physics.msu.ru}

\abstract{First, we briefly review the solar abundance problem and its current status. We start with the description of the existing theoretical model of the Sun and its main uncertainties and proceed with the discussion of the observational evidence for the anomaly. We conclude the review part by outlining the main approaches to solving the solar abundance problem investigated before. In the second part of the paper we consider the poorly studied before non-diffusive energy transport solutions to the problem. We calculate the additional energy flux inside the Sun evidenced by the combined data from helioseismology, spectroscopy and solar neutrino detection experiments. Our findings suggest that the solar abundance problem can be solved by a localized loss of energy near the radiative zone boundary and a gain of approximately the same amount of energy inside the solar core ($0.1R_{\odot} \lesssim r \lesssim 0.3R_{\odot}$). We show that if the localized loss of energy is associated with the resonant emission of transversely polarized hidden photons, the kinetic mixing parameter required does not contradict any of the existing constraints, the ones from the Sun being relaxed as long as the energy loss is compensated by energy deposition inside the core. We introduce then the light dark fermions coupled to hidden photons and discuss their interactions with the solar plasma.  We find the distribution of these dark particles inside the inner part of the radiative zone which is required to provide the necessary heating of the core. }

\keywords{solar physics, stars}

\arxivnumber{}

\begin{document}
\maketitle

\flushbottom

%%%%%%%%%%%%%%%%%%%%%%%%%%%%%%%%%%%%%%%%%%%%%%%%%%%%%%%
%\section{Introduction}\label{sec:mot}
%%%%%%%%%%%%%%%%%%%%%%%%%%%%%%%%%%%%%%%%%%%%%%%%%%%%%%%
\section{Introduction}

Now, half a century after the establishment of the Standard model of particle physics, nobody doubts there is physics beyond it. The main evidence came from the neutrino oscillation experiments, originally from the solar neutrino studies \cite{Ahmad:2002jz}. Compelling reasons for the investigation of different extensions of the Standard model are provided by many astrophysical observations, which favour the existence of yet unknown particles \cite{Zwicky:1933gu, 1996MNRAS.281...27P, Brada__2006, Lage_2014, Aghanim:2018eyx, PhysRevD.98.103005}. In this work we aim to study the discrepancies between our theoretical understanding of the Sun and the observational data, namely the solar abundance problem \cite{2014dapb.book..245B}, assuming they are a possible indication of the physics beyond the Standard model. An important difference with respect to the other similar investigations is that we start with the experimental data and study what general requirements the model must satisfy in order to fit the data. This part of our work has its own value and can be used independently to build the particle physics model which describes the structure of the present-day Sun. In the other part of our article we study if a vector portal extension of the Standard model can provide an example of such a model.
%------------------------------------------------

\section{Solar abundance problem}\label{first}

\subsection{Standard solar model}

Theoretical description of the Sun is based on the Standard solar model \cite{Vinyoles:2016djt}. This model describes evolution of the macroscopic and microscopic solar parameters defined at every point inside the Sun via a system of algebraic and differential equations comprising the equations of hydrostatic equilibrium, continuity, production and transport of energy, thermonuclear reactions, elemental diffusion and the equation of state. One assumes that the Sun is a ball with the fixed mass which was initially chemically homogeneous. Rotation and magnetic fields are not accounted for. The Standard solar model has several unknown parameters that are tuned to fit the well-known values of the solar mass, luminosity, radius, age and surface metallicity. These tuned parameters are initial chemical composition of the Sun and a mixing length -- a phenomenological parameter of the convection model. Apart from them, a result of the modeling contains radial distributions (hereinafter profiles) of luminous flux power, temperature, density, pressure, mass and concentrations of chemical elements inside the present-day Sun, as well as values of solar neutrino fluxes, depending on the parent nuclear reaction. The last thoroughly elaborated Standard solar models are so-called B16 models of the Barcelona group \cite{Vinyoles:2016djt}. Every solar model reproduces the following inner structure of the Sun. In the centre there is a core, the densest and the hottest region of the Sun, where thermonuclear reactions produce power equal to the solar luminosity. The luminous flux corresponding to that power is transported to the outer layers by diffusion through the so-called radiative zone. At some point the temperature gradient falls low enough for plasma to become convectively unstable and the heat transfer becomes convective. Convection zone extends up to the solar surface.

Let $P(r),\, M(r),\, L(r),\, T(r),\, \rho (r) $ be profiles of pressure, mass, luminous flux power, temperature and density inside the Sun, respectively. The main macroscopic equations of the Standard solar model are:
\begin{equation}
\label{SSM}
\left\{ 
\begin{array}{ll}
&\frac{dP}{dr}=-\frac{GM \rho}{r^2},\\[2pt]
&\frac{dM}{dr}=4 \pi r^2 \rho,\\[2pt]
&\frac{dL}{dr} = 4 \pi r^2 \rho \left( \epsilon + \epsilon_{gr} - \epsilon_{\nu} \right) ,\\[2pt]
&\frac{dT}{dr} = \begin{cases} -\frac{3 \kappa \rho L}{16 \pi r^2 a T^3}, \quad \nabla_{rad} < \nabla_{ad} \\[2pt]
\nabla_{ad} \cdot \frac{T}{P} \frac{dP}{dr}, \quad \nabla_{rad} \geq \nabla_{ad}
\end{cases}.
\end{array}
\right.
\end{equation}
where $G$ is a gravitational constant; $a = 4\sigma$, where $\sigma$ is the Stefan-Boltzmann constant; $\nabla_{ad} = \left( \frac{\partial \ln T}{\partial \ln P} \right) _S$ is the adiabatic gradient; $\nabla_{rad} \equiv \frac{3\kappa LP}{16 \pi a GMT^4}$. The energy production rate $\epsilon_{\left( gr,\, \nu \right)}$ and opacity $\kappa$ are functions of density, temperature and chemical composition. The function $\epsilon$ defines energy generation in thermonuclear reactions,  $\epsilon_{\nu}$ -- loss of energy due to the emission of neutrinos, $\epsilon_{gr}$ -- a small addition to the energy generation rate due to the core contraction resulting from the increase of helium abundance. 
The equation of heat transfer has a form of the Fick's equation in the radiative zone, where an absolute value of the temperature gradient is less than the adiabatic gradient. Inside the convection zone the temperature gradient is precisely adiabatic, apart from the very surface layers where the convective flows have a complicated structure and cannot be described by a simple one-dimensional mixing-length theory. One also has to note that besides outer layers of the convection zone, the equation of state of the solar plasma can be accurately approximated by the ideal gas law: $P = \frac{\rho T}{\mu m_H}$, where $m_H$ is the proton mass, $\mu$ is a mean molecular weight equal to the average mass of a particle species inside the mixture normalized to the proton mass; Boltzmann constant is set equal to unity. The chemical composition is parameterized in terms of mass fractions $X_j$ of the chemical elements in a mixture: $X = X_H$ is a fraction of hydrogen, $Y = X_{He}$ corresponds to the fraction of helium, $Z$ -- to the fraction of metals, where all elements heavier than helium are called metals. Using the definition of the mean molecular weight one can easily express it in terms of the mass fractions in case of a fully ionized gas:
\begin{equation}
\label{11}
\mu = \frac{\sum_j n_j A_j}{\sum_j n_j \left( 1+Z_j \right)} = \left[ \sum_j \frac{X_j}{A_j} \left( 1+ Z_j \right) \right] ^{-1} \simeq \left[ 2X + \frac{3Y}{4} + \frac{Z}{2} \right] ^{-1},
\end{equation}
where $A_j$ and $Z_j$ are the atomic mass and charge corresponding to the j-th element; we used an approximation $\left( 1 + Z_j \right) / A_j \simeq 1/2$ for metals.

\subsection{Uncertainties of the theoretical model}\label{errteor}

While building the Standard solar model, the most difficult part is to account properly for microphysical processes, namely to calculate S-factors \cite{Adelberger:2010qa} of various thermonuclear reactions of the proton-proton and CNO cycles, as well as to find an opacity function of the solar plasma. A calculation of the S-factors is a problem of nuclear physics that does not have a good analytical solution in the energy interval required, for one has to deal with the strong coupling regime of quantum chromodynamics. Moreover, parameters of the solar plasma are such that an interaction between two nuclei is highly unlikely and therefore it is not possible to study this process under the solar conditions in a laboratory. Nevertheless, these difficulties were partly overcome with the advantage of effective theories of strong interactions united with an extrapolation of the experimental data obtained for high energies \cite{Adelberger:2010qa, Acharya:2016kfl, Zhang:2018qhm, Marta:2008gg}. As a result, errors of the S-factor values are now quite small so that they do not influence uncertainties of the theoretical determination of those thermodynamic characteristics of the Sun \cite{Vinyoles:2016djt} which are probed by helioseismology. The inaccuracy in the S-factor values is still a dominant source of error while calculating solar neutrino fluxes.

Another important error source is the opacity function of the solar plasma. In order to calculate it, one has to account for a vast of possible atomic transitions, essential contributions coming from light as well as heavy chemical elements, despite the tiny concentration of latter in the plasma. During the last 20 years, cumbersome numerical calculations of the opacity were performed separately by several groups: OP \cite{Badnell:2004rz}, OPLIB \cite{2016ApJ...817..116C}, OPAL \cite{1996ApJ...464..943I}. The opacities obtained are in agreement within an accuracy of several per cent. Besides, there was an experimental measurement of the opacity of iron at the temperature corresponding to the outer layers of the solar radiative zone \cite{Bailey2016}. At these temperatures an iron component of the solar plasma comprises about one fourth of the total opacity. The measured value turned out to be $7\% \pm 4\%$ higher than the theoretically predicted value, that is why the B16 Standard solar models assume that possible errors in the opacity profile can reach 7\%. Recent measurements of chromium and nickel opacities at solar temperatures \cite{PhysRevLett.122.235001} showed smaller disagreements with the theory than in the case of iron. Apart from temperature and density, the arguments of the opacity function include concentrations of heavy elements. The opacity function is very sensitive to plasma metallicity, that is why errors in the determination of metal abundances lead to opacity errors. The total input opacity error is equal to the sum of the error arising from the opacity function calculation and the error arising from the determination of metallicity.

 An opacity profile determines thermal stratification of a given star. Thus, opacity errors are the input data of the Standard solar models which become prime determinants of errors of solar thermodynamic characteristics, namely the sound speed profile and depth of the convection zone. There is also a less trivial fact that the opacity profile greatly influences another output parameter of the Standard solar model -- the surface helium abundance. Let us remind this fact by taking advantage of the equations (\ref{SSM}). The equations of state and hydrodynamic equilibrium give:
\begin{equation}
T \sim \frac{\mu P}{\rho} \sim \frac{\mu G M}{R}.
\end{equation}
Now let us substitute this expression for the temperature together with the continuity equation $\rho \sim M/R^3$ into the Fick's law:
\begin{equation}
\label{2}
L \sim \frac{R^2}{\kappa \rho} \left( \frac{T^4}{R} \right) \sim \frac{\mu^4 M^3}{\kappa}.
\end{equation}
Finally, we use $X\!+Y\!+Z =\!1$, $\,Z \ll X, Y$ and the expression for the mean molecular weight~(\ref{11}):
\begin{equation}
\label{3}
\mu = \frac{1}{2 - 5/4\, Y - 3/2\, Z} \simeq \frac{4}{8-5Y} \quad \Longrightarrow \quad \frac{M^3}{L} \sim \kappa \left( 8-5Y \right)^4.
\end{equation}
Values of the solar luminosity and mass are fixed, thus an opacity change must be compensated by a change in helium abundance. The helium abundance outside the core of the present-day Sun can be changed by tuning the initial helium abundance parameter. The change of the latter parameter determines the change of the solar surface helium abundance, given that variations of elemental diffusion rates play a subdominant role \cite{2010ApJ...714..944V}. Thus, theoretical errors of the calculated surface helium abundance, convection zone depth and sound speed profile are determined mainly by errors of input opacities, comprising an inaccuracy in the opacity function calculation and an inaccuracy in the metallicity determination.

\subsection{Solar abundance problem}

One of the input parameters of the Standard solar model is a ratio of surface abundance of metals to the one of hydrogen $(Z/X)_{\odot} $. The surface abundance of metals is determined by several methods, including spectroscopic analysis of the solar photosphere and corona, chemical analysis of meteorites, as well as theoretical modeling of nucleosynthesis with the use of experimental data on neutron capture cross-sections. Each method is good for some limited range of elements. Meteorite analysis allows one to precisely determine relative fractions of refractory metals \cite{2009LanB...4B..712L}. In order to link these fractions to the abundances of other elements, one calculates abundance of Si independently with two different methods: both chemical analysis of meteorites and spectroscopy of the solar atmosphere. Thus, ratios of the refractory metal abundances to the hydrogen abundance depend on the analysis of Si lines in the spectrum of the photosphere. Abundances of light elements Li, B, Be, C, N, O are determined directly from analysis of the solar spectrum.

In order to accurately determine abundance of a chemical element in the solar photosphere by its spectrum, one needs a good model of the solar atmosphere. Back in the very end of the 20th century the solar spectrum analysis was based on one-dimensional hydrostatic models of the atmosphere, with the assumption of thermodynamic equilibrium. The corresponding solar surface composition GS98 \cite{1998SSRv...85..161G} was later significantly revised due to the development of realistic 3-dimensional non-equilibrium atmospheric models. Besides, previous works were shown to improperly identify an important oxygen line, leading to an overestimate of oxygen abundance \cite{2001ApJ...556L..63A}. The new solar chemical composition AGSS09 \cite{2009ARA&A..47..481A} differs from its predecessor by a significant  metallicity decrease of the solar surface. Physical reasons of such a decrease were well-understood \cite{Grevesse2011}. New atmospheric models allowed one to reconcile the abundances of C, N, O calculated using separately atomic and molecular lines. Nevertheless, despite all the advantages of the chemical composition AGSS09 its decreased metallicity turned out to be a disaster for the Standard solar model. Before going into details let us get away from the solar surface chemical composition and briefly discuss another important source of knowledge about the Sun -- helioseismology \cite{2016LRSP...13....2B}. Sun and other stars oscillate with the frequencies determined by their normal modes. These oscillations are caused by motions of plasma in the convection zone and have very small amplitudes, so that they can be described as linear adiabatic oscillations. Adiabaticity is violated in the solar surface layer, but the corresponding errors can be eliminated during the analysis. The waves corresponding to the oscillations of different modes can penetrate deep inside the Sun, thus allowing one to precisely determine the inner characteristics of the Sun as soon as the frequencies of oscillations at the surface are measured. In order to measure the solar oscillations precisely one needs a net of telescopes. In particular, important data was obtained with BiSON net \cite{2007ApJ...659.1749C} and MDI experiment \cite{1995ASPC...76..402S} on the board of SoHO observatory \cite{1995SoPh..162....1D}. The frequency spectra obtained are subject to the procedure of helioseismological inversion, involving the Fourier transform of the spectra with the help of linearization of solar profiles near the theoretical profiles of some reference Standard solar model. For the scope of our analysis we will use inversions of the BiSON net results \cite{Basu:2009mi}.

Helioseismology allowed one to determine some characteristics of the Sun, such as sound speed and density profiles, depth of the convection zone and surface helium abundance, with a relative accuracy of several per mille or better. The relative accuracy of the theoretical prediction of these parameters in the frame of the B16 Standard solar models is somewhat lower, though it is also about several per mille. This allows one to make a constructive comparison between the theoretical calculations and the helioseismological results. It turns out that the helioseismological data do not agree with the Standard solar model, overall significance of the discrepancy being $4.7\sigma$ \cite{Vinyoles:2016djt}. The discrepancy can be diminished by switching to the older high-metallicity chemical composition GS98 which yields the anomaly significance of $2.7\sigma$. Due to the dependence of the discrepancy on surface chemical composition, the corresponding breach in the Standard solar models was called the solar abundance problem. The main source of the anomaly is a discrepancy between theoretical and helioseismological sound speed profiles in the solar radiative zone. Besides, discrepancies of less significance are found between the theoretical and helioseismological determinations of depth of the convection zone as well as surface helium abundance. One has to mention that the solar abundance problem concerns not only solar physics, but also astrophysics as a whole. The methods and theories which are used to describe the Sun are quite general and determine physics of any other normal star, that is why a breach in the Standard solar model may lead to misinterpretation of observational data from other stars and stellar populations.

\subsection{Studies of the solar abundance problem}\label{solutions}

The revision of chemical composition of the solar surface and subsequent emergence of the solar abundance problem motivated a lot of theoretical studies aiming to solve the problem, but none of them has succeeded to provide a physically justified non-contradictory solution. The first attempt at the solution we mention is to increase the opacity function in the solar radiative zone. As it was mentioned earlier in the section \ref{errteor}, a variation of opacity changes predictions of the solar model about the parameters which are known from helioseismology: the sound speed profile, depth of the convection zone and surface helium abundance. Lower metallicity of the solar model with the AGSS09 chemical composition makes the solar plasma less opaque. Since models with the high metallicity GS98 are consistent with the helioseismology within $3\,\sigma$, it is clear that a solution to the solar abundance problem can be found by increasing the opacity function: a change in opacity due to the metallicity variation can be compensated by a change in values of the opacity function themselves. Using the Standard solar model numerical framework one can determine the opacity profile which corresponds to the helioseismological data \cite{2010ApJ...724...98V, Song:2017kvf, Villante:2015xla}. It turns out that in order to reproduce such a profile in the models with the chemical composition AGSS09, one has to increase the opacity function by 20-30\%. As it was mentioned in the section \ref{errteor}, a comparison of the different theoretical calculations of the opacity function between themselves, as well as with the existing experimental measurements suggests much lower errors, not higher than about 7\%. Thus, while an opacity increase could have been a straightforward solution to the solar abundance problem, the data we have do not allow us to consider it a viable one.

Another branch of investigations of the solar abundance problem comprises studies of gravitational settling of metals inside the Sun: heavier elements settle towards the solar centre increasing a gradient of metal abundance. If one could enhance gravitational settling of metals in the solar models, the low surface metallicity would be explained by a higher rate of escape of metals from the surface. Meanwhile, metallicity inside the Sun would be higher. It was figured out that the required enhancement of the gravitational settling is too high to find a viable physical justification for it. Besides, this hypothetical scenario turned out to provide only a partial solution to the problem \cite{Guzik:2005db, Guzik:2010ck}. Although there has been recently a work where the authors show that after taking advantage of the solar rotation, this scenario can in principle reproduce the helioseismological data \cite{2019ApJ...873...18Y}, the physical justification for the main ingredient -- enhancement of the gravitational settling -- is still lacking. Thus, these attempts at the solution have the same status as the ones discussed in the previous paragraph.

Many other investigations of the solar abundance problem, such as a study of influence of accretion on the solar models \cite{Serenelli:2011py}, a revision of Ne abundance \cite{Bahcall:2005dd}, attempts at building the general non-standard solar models \cite{2010ApJ...715.1539T} and models with the surface chemical composition evidenced by solar wind studies \cite{2016MNRAS.463....2S, 2017ApJ...839...55V}, as well as others, were not able to resolve the problem. The inability of the well-established physics to explain the anomaly gave impetus to studies of the problem in the context of models beyond the Standard model of particle physics. In particular, authors of the work \cite{Vincent:2012bw} considered a possible influence of hypothetical light particles (axion-like particles, chameleons and paraphotons) on the formation of absorption spectra in the photosphere: it was assumed that the spectroscopic data concerning the surface composition could be wrong due to a spectrum distortion in the presence of these hypothetical particles. In the work \cite{Zanzi:2014aia} authors investigated an influence that a chameleon field could exhibit on the solar observables. Another investigation attempted at reconciliation of the solar model with helioseismological data involves hypothetical dark matter particles with a velocity or momentum-dependent cross-section of interaction with the ordinary matter \cite{Vincent:2015gqa, Geytenbeek:2016nfg}. Presence of these particles inside the core could influence energy balance inside the Sun and therefore partially reconciliate the solar models with helioseismological data. However, required dipole moments of the dark matter particles are well excluded by different dark matter search experiments. Moreover, the masses required are quite small, so that an effect of evaporation of the dark matter become important, while the authors do not account for this effect. None of the investigations discussed have succeeded to solve the solar abundance problem.

%------------------------------------------------

\section{Non-diffusive energy transport solution}\label{second}

\subsection{Problem statement}\label{task}

One of the easiest among the proposed solutions of the solar abundance problem is to increase the opacity profile in the solar radiative zone, though the physical mechanism which could lead to such an increase is not known. From the point of solar physics such an increase leads to a change of the thermal stratification of a star due to a variation of energy transfer parameters. Let us generalize this approach to the resolution of the solar abundance problem and wonder what energy transfer change is required to solve the problem. 

In the particular case of diffusive energy transport the answer is already known: as it is discussed in the section \ref{solutions}, one needs a definite change of opacity in the radiative zone, which was calculated in the work \cite{Villante:2015xla}. Let us also mention that if one is interested in the only one helioseismological parameter, namely the sound speed profile, which is the main source of the anomaly, the requirements on the opacity change can be relaxed \cite{2010ApJ...724...98V}. Global rescaling of opacity does not influence the sound speed profile \footnotemark, that is why from the point of the sound speed profile, an opacity increase in the radiative zone is equivalent to an opacity decrease in the core. \footnotetext{One can show that this is true by using the virial theorem -- the total internal energy of a star is proportional to the total gravitational energy: $\int\! dm\, Gm/r = 2 \cdot 3/2 \int P dV = 3 \int \! dm\, P/\rho \propto \int \! dm\, c_s^2$. The present-day solar radius is fixed in the solar model, so after a global rescaling of the opacity the left part does not change. Thus, the sound speed $c_s$ does not vary as well. This argument was also checked numerically in the work \cite{2010ApJ...724...98V}.} A less opaque core can be constructed by establishing a new diffusive energy transport inside the core. This is what the authors of the works \cite{Vincent:2015gqa, Geytenbeek:2016nfg} use, trying to reproduce the helioseismological sound speed profile by putting appropriate dark matter inside the solar core. As it is obvious from the expression (\ref{3}), an opacity decrease inside the core leads to a decrease of the surface helium abundance. The Standard solar model predicts an underestimated value for this parameter compared to the helioseismological value \cite{Vinyoles:2016djt} even without this additional decrease, thus additional diffusive energy transport inside the core can provide only a partial solution to the solar abundance problem. Nevertheless, statistically, this partial solution turns out not to be completely unacceptable: the statistical analysis presented in the work  \cite{Geytenbeek:2016nfg}, which discusses additional energy transport inside the core due to dark matter particles, showed that some of the models studied can reproduce helioseismological data with the probability about several per cent. This means that the required opacity increase, calculated in the work \cite{Villante:2015xla}, can be complemented by a small global opacity rescaling. If the rescaling is small enough, agreement between the corresponding solar model and helioseismological data continues to be good.

Now, let us consider generic non-diffusive energy transport, which has not been studied earlier in the context of solar physics. By the non-diffusive energy transport here we mean an analogue of radiative transport, because convective energy transport inside the solar core or radiative zone (not including the possible thin overshoot layer) is obviously not consistent with helioseismological data. This analogue of radiative transport can be implemented via emission of some unknown particles from one region of the Sun, their propagation and/or capture inside the Sun and subsequent energy transfer from them to the solar matter within another region of the Sun. In the context of the solar equations, generic non-diffusive energy transport can be parameterized by a loss of energy from one region of the Sun and its deposition into another. Thus, additional terms will emerge in the right-hand side of the energy production equation of the system (\ref{SSM}). This equation already contains a term characterizing emission of weakly interacting particles, namely neutrinos. Emission ($\epsilon_i < 0$) of other practically non-interacting with matter particles from the solar core and its influence on the solar physics were studied in many works, for example \cite{Schlattl:1998fz, Gondolo:2008dd, Vinyoles:2015aba, Vinyoles:2015khy}. In our case, we will allow terms $\epsilon_i$ to have any possible sign and will determine what energy loss/deposition profile $\epsilon$ is evidenced by combined data from helioseismology, present-day techniques of determination of chemical composition of the solar surface and theoretical opacity calculations. In order to do that, we will calculate an evidenced profile of power of the luminous flux, which is tightly connected to the profile of energy loss/deposition.

\subsection{Optimal profile of luminous flux power}\label{opt}
\subsubsection{Main equations, conventions and boundary conditions}
In order to determine the optimal energy loss/deposition profile, let us first consider the solar equations (\ref{SSM}) and perform a change of the independent variable from the radial coordinate to the mass one. The mass coordinate is normalized by the solar mass $m \equiv M/M_{\odot}$. The solar mass is fixed, so the mass coordinate lies in the interval $ [0,1]$. For the sake of convenience, let us rescale other variables of the equations (\ref{SSM}) as well. We define $\tilde{r} \equiv r/R_{\odot}$, $\, \tilde{\rho} \equiv \rho/\rho_0$, $\, \tilde{p} \equiv P/P_0$, $\, \tilde{l} \equiv L/L_{\odot}$, $\, \tilde{t} \equiv T/T_0$, $\, \tilde{\kappa} \equiv \kappa / \kappa_0$, $\, \tilde{\epsilon} \equiv \epsilon/\epsilon_0$, where
\begin{equation*}
\rho_0 \equiv \frac{M_{\odot}}{\frac{4}{3} \pi R_{\odot}^3}, \quad P_0 \equiv \frac{2}{3} \pi R_{\odot}^2 G \rho_0^2, \quad T_0 \equiv \frac{GM_{\odot}m_H}{R_{\odot}}, \quad \kappa_0 \equiv 1.0\,\, \frac{\text{cm}^2}{\text{g}}, \quad \epsilon_0 \equiv 1.0\,\, \frac{\text{erg}}{\text{g}\cdot \text{s}}.
\end{equation*}
We remind that the Boltzmann constant is set equal to unity. We will omit tilde above the letters, assuming that all the variables are by default rescaled. In order to reexpress the system (\ref{SSM}) in terms of the mass coordinate, we divide all its equations by the second equation and calculate a derivative of the radial coordinate with respect to the mass coordinate using the inverse derivative rule applied to the continuity equation. The resulting solar equations are:
\begin{itemize}
	\item the hydrostatic equilibrium equation:
	\begin{equation}\label{hydr}
	p' = -\frac{2 m}{3 r^4},
	\end{equation}
	\item the continuity equation:
	\begin{equation}\label{contin}
	r' = \frac{1}{3 \rho \, r^2},
	\end{equation}
	\item the energy production equation:
	\begin{equation}\label{energy}
	l' = \xi_1 \cdot \left( \epsilon_{nuc} - \epsilon_{\nu} + \sum_i \epsilon_i \right),
	\end{equation}
	\item the energy transport equation:
	\begin{equation}\label{fick}
	t' = -\xi_2 \cdot \frac{\kappa \, l}{r^4\, t^3},
	\end{equation}
	\item the equation of state:
	\begin{equation}\label{state}
	p = 2 \cdot \frac{\rho \, t}{\mu},
	\end{equation}
\end{itemize}
where the numerical coefficients $\xi_1$ and $\xi_2$ are defined as following:
\begin{equation}\label{coeff}
\xi_1 \equiv \epsilon_0 \cdot \frac{M_{\odot}}{L_{\odot}} \approx 0.52, \quad \xi_2 \equiv \frac{3}{ 64 \pi^2 a} \cdot \frac{\kappa_0 \, L_{\odot} M_{\odot}}{\left( G M_{\odot} m_H \right) ^4} \approx 2.4 \cdot 10^{-5}.
\end{equation}
We neglected the energy deposition $\epsilon_{gr}$, which originates from gravitational restructuring of the Sun, because the present-day Sun practically does not change its structure. Besides, we added the terms $\epsilon_i$, which parameterize additional non-diffusive energy transport.

Now, let us consider the energy production equation (\ref{energy}) and integrate it in order to find an expression for the luminous flux power:
\begin{equation}\label{fluxpower}
l \left( m \right) = \xi_1 \int\limits_0^m \left( \epsilon_{nuc} \left( \bar{m} \right) - \epsilon_{\nu} \left( \bar{m} \right) + \sum_i \epsilon_i \left( \bar{m} \right) \right) d\bar{m}.
\end{equation}
We note that the luminous flux is equal to zero $l \left( 0 \right) = 0$ in the solar centre, which can be easily understood by symmetry considerations as well. There is another boundary condition on the solar surface where $l \left( 1 \right) = 1$, for the luminous flux power on the surface must be equal to the well-known value of the solar luminosity. In the framework of the Standard solar models there is no term $ \sum_i \epsilon_i $ in the integrand of the equation (\ref{fluxpower}), that is why the boundary condition on the surface can be written as following:
\begin{equation}
\xi_1 \int\limits_0^1 \left[ \epsilon_{nuc} \left( m \right) - \epsilon_{\nu} \left( m \right) \right]_{\text{SSM}} dm = 1,
\end{equation}
where the subscript 'SSM' denotes that a function is calculated within the framework of the Standard solar model. The integrand is determined by a rate of thermonuclear reactions inside the Sun. If $\int_0^1 dm \sum_i \epsilon_i \neq 0$, the boundary condition on the surface requires a change of the rate of these reactions: $ \epsilon_{nuc} - \epsilon_{\nu} \neq \left( \epsilon_{nuc} - \epsilon_{\nu} \right) _{\text{SSM}} $. Let us express an arbitrary addition to the energy loss/deposition profile in the following way:
\begin{equation}
\sum_i \epsilon_i = \sum_i \tilde{\epsilon}_i \, + \, \bar{\epsilon}\, , \quad \bar{\epsilon}\, \equiv \, \int\limits_0^1 dm \sum_i \epsilon_i.
\end{equation}
Then $\int_0^1 dm \sum_i \tilde{\epsilon}_i = 0\,$: we have separated the part corresponding to the non-diffusive energy transfer itself. The value $\bar{\epsilon}$ determines the total power deposited into the solar plasma by unknown sources or lost by the plasma through unknown energy sinks. Output parameters of the solar models which are sensitive to $\bar{\epsilon}$ are the values of neutrino fluxes, because the total power of the solar thermonuclear reactor changes by $\bar{\epsilon}$. The function $\sum_i \tilde{\epsilon}_i \left( m \right)$, which determines the variation of the luminous flux power profile $l \left( m \right)$, does not influence the total power of the thermonuclear reactor. In order to find the required additional energy transfer $\sum_i \tilde{\epsilon}_i \left( m \right)$, let us calculate that variation of $l \left( m \right) $ with respect to the B16 Standard solar model, which is required to solve the solar abundance problem.

\subsubsection{Connection to helioseismological data}

Let us consider two solar models -- the Standard solar model and a model with the additional energy transfer $\sum_i \tilde{\epsilon}_i \left( m \right)$, which solves the solar abundance problem. All quantities of the second model will be equipped with a tilde, while all quantities of the first one will not. We denote $\: \delta x \equiv \tilde{x} - x$, which quantifies a difference in the value of a parameter $x$ between the two models. For a finite change of a product of two quantities $a\cdot b \propto c$ one can write:
\begin{equation*}
\delta \left( a \cdot b \right) = \tilde{a}\cdot \tilde{b} - a \cdot b = \delta a \cdot b + a \cdot \delta b + \delta a \cdot \delta b \quad \Rightarrow \quad \frac{\delta c}{c} = \frac{\delta a}{a} + \frac{\delta b}{b} + \frac{\delta a}{a} \cdot \frac{\delta b}{b},
\end{equation*}
independently of the proportionality coefficient. This expression is valid in case $a, b \neq 0$. If the relative change of $a$ or $b$ is small compared to unity, the last term can be neglected. In the general case of a product of several quantities, cross-products of only those relative changes are significant, which are comparable to unity . Quite analogously, for a ratio of two quantities $a/b \propto g$ one can write the following:
\begin{equation*}
\frac{\delta g}{g} = \frac{\delta a}{a} - \frac{\delta b}{b} \cdot \frac{1}{1+\delta b /b} - \frac{\delta a}{a} \cdot \frac{\delta b / b}{1+ \delta b /b}.
\end{equation*}

\begin{figure}[h!]
	\begin{floatrow}
		\ffigbox{\caption{Opacity coefficient $\alpha$ as a function of mass coordinate}\label{opal_a}}
		{\includegraphics[height=5cm, width=7cm]{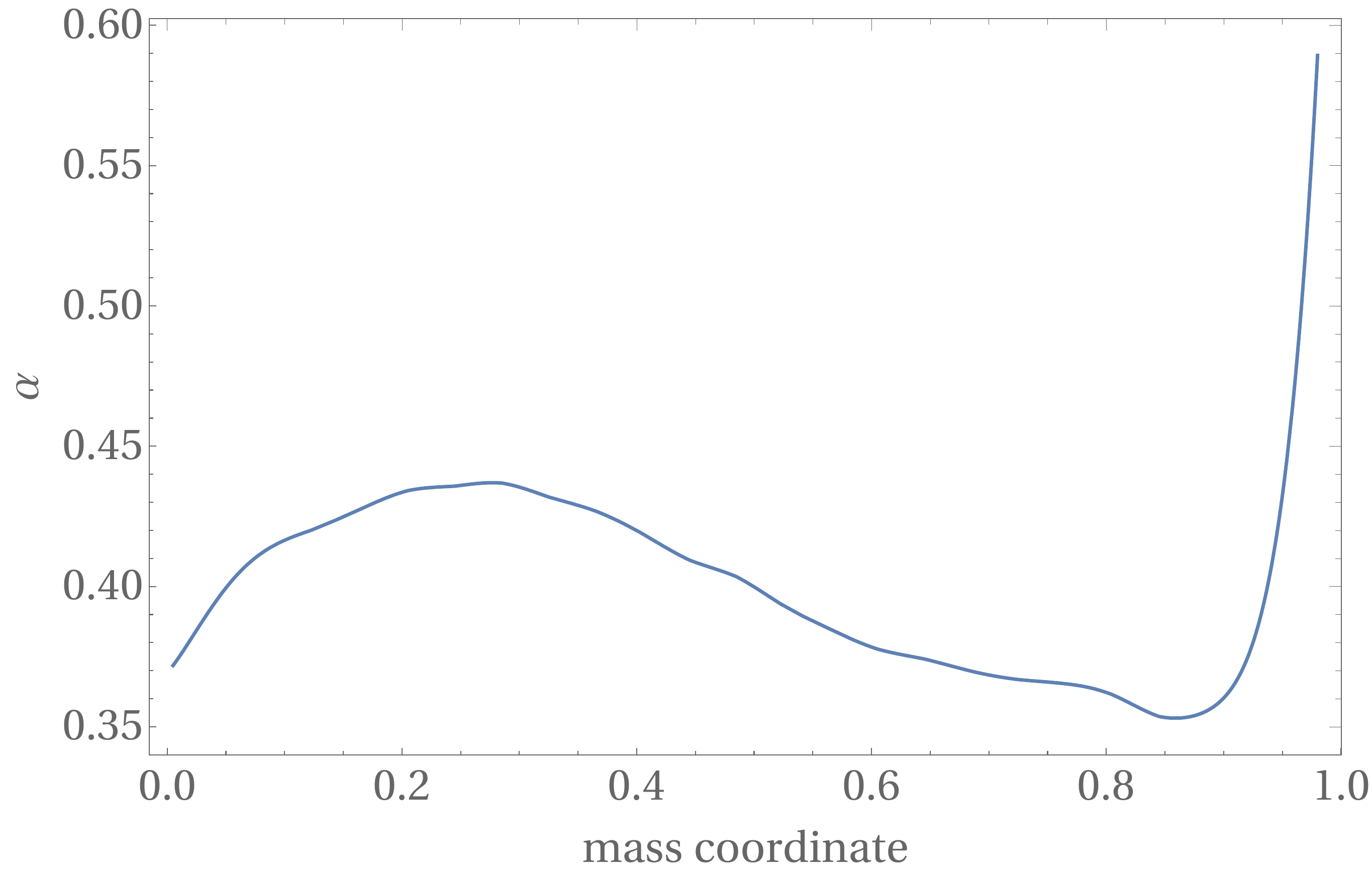}}
		\ffigbox{\caption{Opacity coefficient $\beta$ as a function of mass coordinate}\label{opal_b}}
		{\includegraphics[height=5cm, width=7cm]{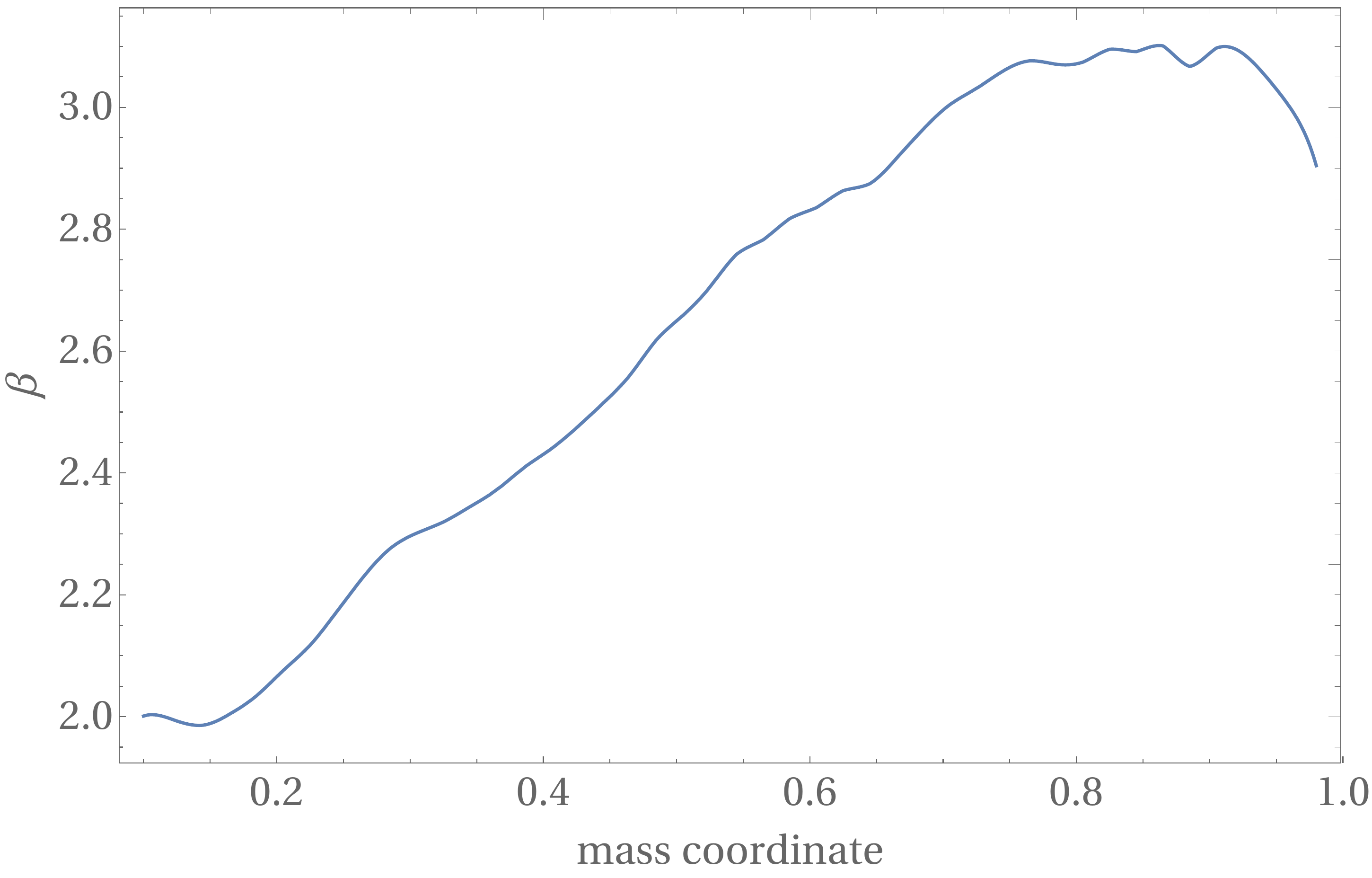}}
	\end{floatrow}
\end{figure}

\begin{figure}[h!]
	\begin{floatrow}
		\ffigbox{\caption{Relative sound speed variation required to reconcile the B16 Standard solar model with helioseismology data as a function of radial coordinate; surface chemical composition AGSS09}\label{deltacs}}
		{\includegraphics[height=5cm, width=7cm]{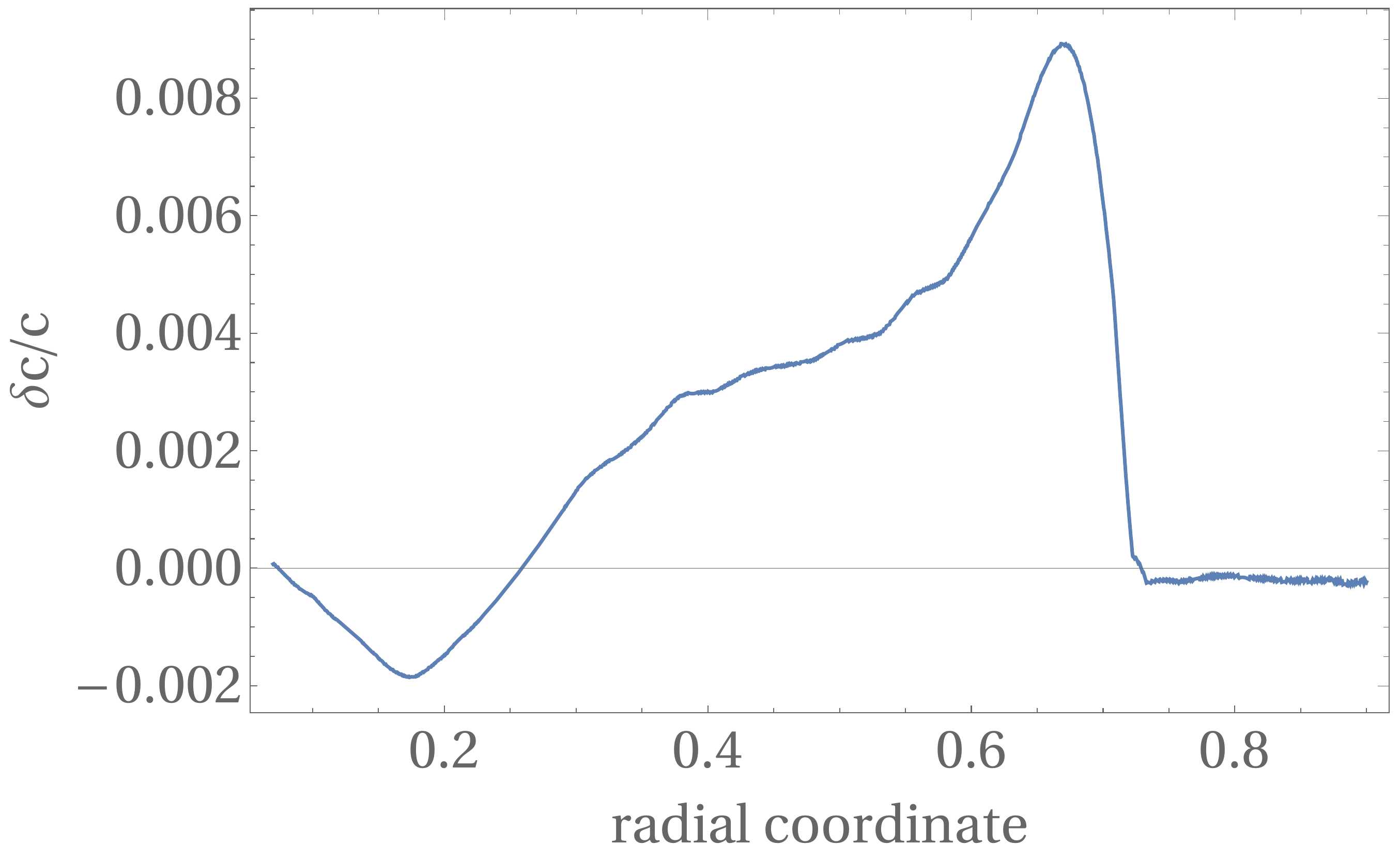}}
		\ffigbox{\caption{Relative density variation required to reconcile the B16 Standard solar model with helioseismology data as a function of radial coordinate; surface chemical composition AGSS09}\label{deltaro}}
		{\includegraphics[height=5cm, width=7cm]{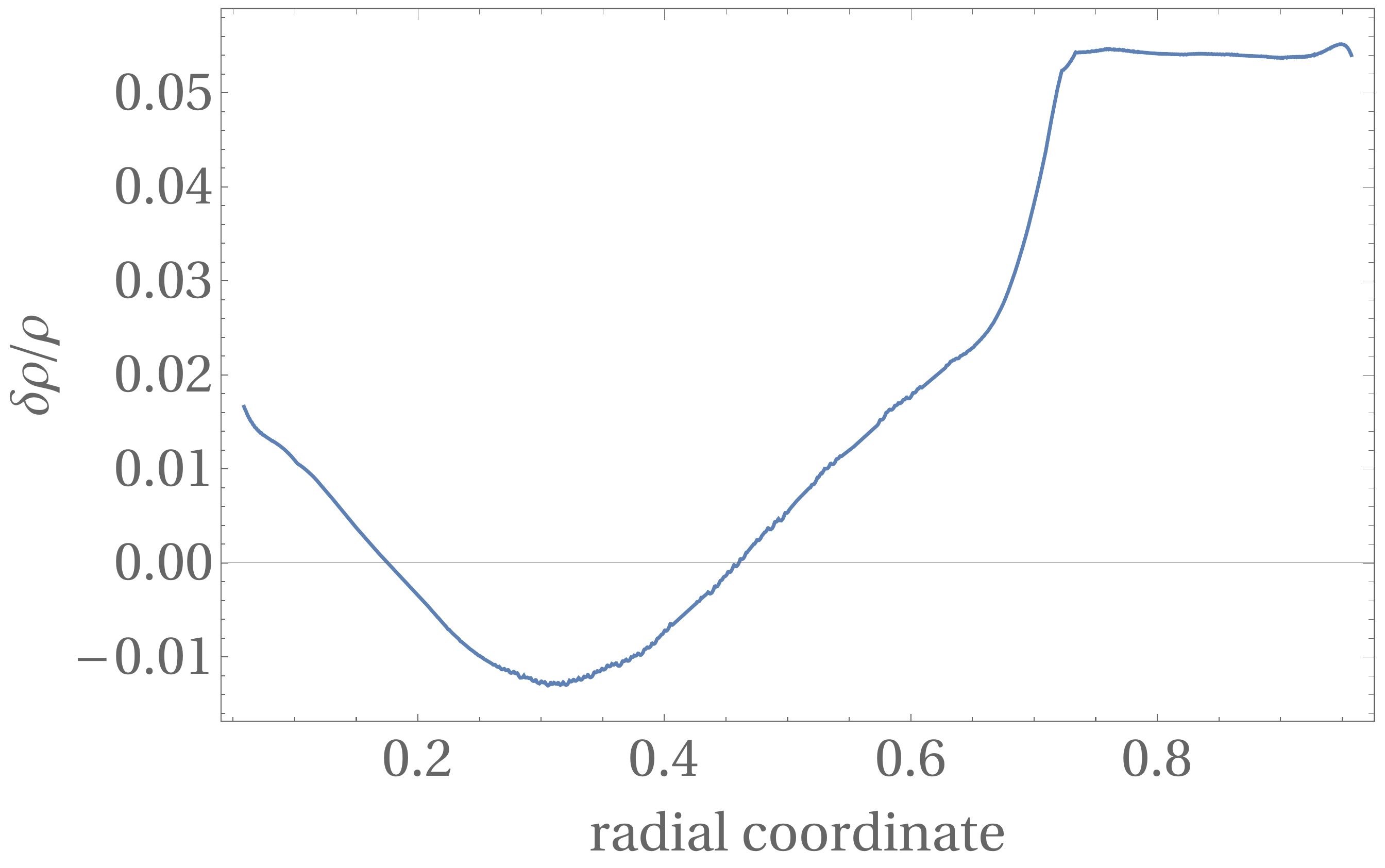}}
	\end{floatrow}
\end{figure}

In order to calculate the required variation of the luminous flux power profile $l \left( m \right)$, let us use the Fick's law (\ref{fick}) and substitute the expression for the radial coordinate according to the hydrostatic equilibrium equation (\ref{hydr}). As a result, we get an expression which is valid at each point $m$:
\begin{equation}\label{begin}
l \propto \frac{m\, t^3\, t'}{\kappa \, p'}.
\end{equation}
Opacity is a function of density, temperature and metallicity. In the models we consider, the latter is fixed by the solar composition AGSS09. A dependence of the opacity on density and temperature at each point $m$ can be parameterized by two coefficients $\alpha$ and $\beta\:$: $\, \kappa \propto \rho^{\alpha} \, t^{-\beta}$. For example, in the case of the Kramer's law, which is violated inside the Sun due to a significant contribution to the opacity from bound-bound atomic transitions, these coefficients have the well-known values $\alpha = 1, \; \beta = 7/2$. We use the OPAL opacity tables \cite{1996ApJ...464..943I} as well as interpolation routines \cite{opal_interpolation} in our calculations. We find an opacity function for plasma with the chemical composition AGSS09 and the helium abundance, which is determined by the helium abundance profile of the B16 Standard solar model \cite{Vinyoles:2016djt}. The calculation of the coefficients $\alpha$ and $\beta$ using the fixed helium profile is justified by the fact, that these coefficients only weakly depend on helium abundance, which is almost the same in the two different models under consideration, as we show later. The opacity function is taken in 196 points from $m = 0.005$ till $m = 0.980$ -- all the way through the radiative zone. After interpolation, we obtain the function $o = \log_{10} \kappa \left( v, w \right) \,$, where $v = \log_{10} T, \: w = \log_{10} \rho / T_6^3, \: T_6 \equiv T/10^6 \text{K}$. Then the coefficients $\alpha$ and $\beta$ are calculated:
\begin{equation}\label{albe}
\alpha = \frac{\partial \log_{10} \kappa}{\partial \log_{10} \rho} = \frac{\partial o}{\partial w}, \quad \beta = - \frac{\partial \log_{10} \kappa}{\partial \log_{10} T} = 3\cdot \frac{\partial o}{\partial w} - \frac{\partial o}{\partial v}.
\end{equation}
Their dependence on the mass coordinate is given in Fig. \ref{opal_a} and \ref{opal_b}.
Derivatives of the function $o\left( v, w \right)$ were calculated at each point, at temperature and density corresponding to the B16 Standard solar model. This is again justified by the fact that a difference in the temperature and density values between the two models under consideration is small. Let us show this smallness by taking advantage of the helioseismological data \cite{Basu:2009mi}, in particular of the data on sound speed and density profiles as well as on surface abundance of helium. Helioseismologically required changes in sound speed and density with respect to the B16 Standard solar model never exceed the level of several per cent, as it can be seen from Fig. \ref{deltacs} and \ref{deltaro}.
Temperature is proportional to sound speed squared $c_s^2 \propto p/\rho \propto t/\mu$, that is why having accounted for the smallness of the sound speed variation, one can write:
\begin{equation}\label{deltat}
\frac{\delta t}{t} = 2\, \frac{\delta c_s}{c_s} + \frac{\delta \mu}{\mu}.
\end{equation}
According to the formula (\ref{3}), mean molecular weight is determined by helium abundance. Given the surface helium abundance in the B16 Standard solar model $Y = 0.23$ and its helioseismological value $\tilde{Y} = 0.25$, one can calculate the relative variation of the mean molecular weight: $\delta \mu / \mu \approx 0.01$. A significant variation of helium abundance in the centre of the Sun would lead to a large variation of rates of thermonuclear reactions, which is possible only if $\bar{\epsilon} \neq 0$. This scenario is constrained by solar neutrino flux data. In particular, results of the work \cite{Gondolo:2008dd} suggest that $l_{\text{loss}} < 0.1 \: \Rightarrow \: |\bar{\epsilon}\, | < 0.1/\xi_1 = 0.2$. In the Standard solar model, the amount of helium produced by thermonuclear reactions is $ Y_{\text{nuc}} = 0.34$. This quantity is proportional to the rate of thermonuclear reactions, therefore we can easily calculate its variation: $\delta Y_{\text{nuc}} = Y_{\text{nuc}} \cdot | \bar{\epsilon}\, | < 0.07$. Taking into account that, according to the Standard solar model, the total abundance of helium in the centre of the Sun is $Y_c = 0.62$, we estimate the relative change of the mean molecular weight which originates from a variation of the rate of thermonuclear reactions:
\begin{equation}\label{munuc}
\left( \frac{\delta \mu}{\mu} \right)_{\text{nuc}} = \left( \delta \ln \mu \right)_{\text{nuc}} = \frac{\partial \ln \mu}{\partial Y} \cdot \delta Y_{\text{nuc}} = \frac{5\, \delta Y_{\text{nuc}}}{8 - 5\, Y_c} < 0.07.
\end{equation}
Thus, the considered relative variations of density, sound speed, temperature and mean molecular weight are all much less than unity.

Let us also show that $\left| \delta p'/p' \right| \ll 1$. First, we consider the continuity equation (\ref{contin}). From this equation one can infer that $\rho \left( r^3 \right)' = \text{const}$. Taking into account smallness of the relative variation of density one can write:
\begin{equation}\label{cap}
\frac{\delta \rho}{\rho} + \frac{\delta \! \left( r^3 \right)'}{\left( r^3 \right)'} = 0.
\end{equation} 
The following equalities hold as well:
\begin{equation}
r^3 \left( m \right) = \int\limits_0^m  \left( r^3 \right)' \left( x \right) dx \quad \Rightarrow \quad \delta r^3 \left( m \right) = \int\limits_0^m  \delta \left( r^3 \right)' \left( x \right) dx.
\end{equation}
The last equality is rewritten with the use of (\ref{cap}):
\begin{equation}\label{integ}
\delta r^3 \left( m \right) = - \int\limits_0^m  \frac{\delta \rho}{\rho} \cdot \left( r^3 \right)' \left( x \right) dx.
\end{equation}
The function $\delta \rho / \rho $ is bounded and continuous, while the function $\left( r^3 \right)' $ is not negative by its meaning over any possible interval $[0, m]$. Thus, one can apply the mean value theorem to the integral (\ref{integ}):
\begin{equation}
\int\limits_0^m  \frac{\delta \rho}{\rho} \cdot \left( r^3 \right)' \left( x \right) dx =  \frac{\delta \rho}{\rho} \left( \bar{m} \right) \cdot \int\limits_0^m \left( r^3 \right)' \left( x \right) dx = \frac{\delta \rho}{\rho} \left( \bar{m} \right) \cdot r^3 \left( m \right),
\end{equation}
where $\bar{m} \in [0, m]$. Then $\delta r^3\! \left( m \right) / r^3\! \left( m \right) = - \delta \rho \left( \bar{m} \right) / \rho \left( \bar{m} \right) $, which means that the relative variation $3\, \delta r / r$ is small. Now, let us consider the hydrostatic equilibrium equation (\ref{hydr}), which gives the relation between the pressure derivative and the fourth power of the radial coordinate. Due to smallness of the quantity $r^3$, the quantity $r^4$ is small as well, that is why:
\begin{equation}
\left| \frac{\delta p'}{p'} \right|= 4 \left| \frac{\delta r}{r} \right| \ll 1.
\end{equation}

Now, let us return to the formula (\ref{begin}) and substitute into it an expression for the opacity with the coefficients (\ref{albe}). Taking into account smallness of the relative variations of temperature and density as well as pressure derivative, we get an expression for the relative variation of the luminous flux power:\begin{equation}\label{interm}
l \propto \frac{m \, t^{3+\beta} \, t'}{\rho^{\alpha} \, p'} \quad \Rightarrow \; \frac{\delta l}{l} = \left( 3+ \beta \right) \frac{\delta t}{t} - \alpha \, \frac{\delta \rho}{\rho} + \frac{\delta t'}{t'} - \frac{\delta p'}{p'}.
\end{equation}
The latter equality is valid at any point of the radiative zone and the core, except for the point $m = 0$, where the boundary condition requires $\delta l = 0$. Next, we show that the terms in the right-hand side of the equality (\ref{interm}) can be unambiguously rewritten in terms of helioseismological data and a variation of the chemical composition. We note that
\begin{equation}\label{deltadt}
\frac{\delta t'}{t'} = \frac{\left( \delta t\right)'}{t'} = \frac{t}{t'} \, \left( \frac{\delta t}{t} \right)' + \frac{\delta t}{t} = \frac{2\, t}{t'}\, \left( \frac{\delta c_s}{c_s} \right)' + \frac{t}{t'} \left( \frac{\delta \mu}{\mu} \right)' + 2\, \left( \frac{\delta c_s}{c_s} \right) + \frac{\delta \mu}{\mu},
\end{equation}
where we have taken advantage of the equality (\ref{deltat}). Analogously, due to proportionality $c_s^2 \propto p/\rho$ and smallness of the relative variation of density, one can write:
\begin{equation}\label{deltadp}
\frac{\delta p'}{p'} = \frac{p}{p'} \, \left( \frac{\delta p}{p} \right)' + \frac{\delta p}{p} = \frac{p}{p'}\, \left( \frac{\delta \rho}{\rho} \right)' + 2\, \frac{p}{p'} \left( \frac{\delta c_s}{c_s} \right)' + 2\, \left( \frac{\delta c_s}{c_s} \right) + \frac{\delta \rho}{\rho}.
\end{equation}
As a result, we get the following expression for the variation of the luminous flux power:
\begin{equation}\label{deltal}
\frac{\delta l}{l} = \, 2\cdot \left( 3+ \beta \right) \frac{\delta c_s}{c_s} - \left( 1+ \alpha \right) \frac{\delta \rho}{\rho} + \left( 4+\beta \right) \frac{\delta \mu}{\mu} \, +\frac{t}{t'} \left[ 2\, \left( \frac{\delta c_s}{c_s} \right)' + \left( \frac{\delta \mu}{\mu} \right)' \, \right] - \frac{p}{p'} \left[ \left( \frac{\delta \rho}{\rho} \right)' + 2 \left( \frac{\delta c_s}{c_s} \right)'\,\right] .
\end{equation}
Now, we can see that, hypothetically, if one had accurate helioseismological inversions of sound speed and density as well as a value for the surface helium abundance, then, after having set some value for the parameter $\bar{\epsilon}$, it would be possible to calculate $\delta c_s / c_s$, $\delta \rho / \rho$ and $\delta \mu / \mu $ profiles and finally determine the required $\delta l / l$ profile according to the formula (\ref{deltal}). Nevertheless, the helioseismological inversions as well as the outputs of the Standard solar model (see section \ref{errteor}) have their errors, so a straightforward calculation of $\delta l / l $ using the formula (\ref{deltal}) does not fix the shape for the required variation of luminous flux power. We will overcome these difficulties by taking advantage of the results of previous investigations of the solar abundance problem, namely of the hypothetical solution \cite{Villante:2015xla} by means of an increase of opacity, which we mentioned in the section \ref{solutions}.

\subsubsection{Method of equivalent opacity variation}

We know that some special kind of opacity change can solve the solar abundance problem. Let us consider two solar models: the first one let again be the B16 Standard solar model, the second one -- a model with the opacity change that solves the solar abundance problem. Analogously to our previous notation for the variation of some quantity from one solar model to another, we denote a difference in $x$ between these two models as $\bar{\delta} x$. Then let us use an expression similar to (\ref{begin}) in order to relate a variation of the opacity to the variations of other quantities:
\begin{equation}\label{beginop}
\kappa \propto \frac{m\, t^3\, t'}{l \, p'} \quad \Rightarrow \quad \frac{\bar{\delta} \kappa}{\kappa} = 3\, \frac{\bar{\delta} t}{t} + \frac{\bar{\delta} t'}{t'} - \frac{\bar{\delta} p'}{p'} - \left( \frac{\bar{\delta} l}{l} \right)_{\text{nuc}}.
\end{equation}
Profiles of the luminous flux power in the two models under consideration differ by a function $\left( \bar{\delta} l \right)_{\text{nuc}}$, which is not zero only within the solar core and corresponds to a change in the thermonuclear fusion profile $\epsilon = \epsilon_{nuc} - \epsilon_{\nu}\,$ due to variations of the profiles of density, temperature and chemical composition inside the core. Let us justify the equality (\ref{beginop}) by showing that the relative variation of power is small in this case. The equation (\ref{energy}) and the mean value theorem give:
\begin{equation}\label{avev}
\left( \bar{\delta} l \right)_{\text{nuc}} \left( m \right) = \xi_1 \int\limits_0^m \epsilon \cdot \frac{\bar{\delta} \epsilon}{\epsilon} \left( x \right) dx = \xi_1 \, \frac{\bar{\delta} \epsilon}{\epsilon} \left( \bar{m} \right) \int\limits_0^m \epsilon \, dx = \frac{\bar{\delta} \epsilon}{\epsilon} \left( \bar{m} \right) \cdot l \left( m \right),
\end{equation}
where $\bar{m} \in [0, m]$. Mean value theorem is valid, for the function $ \bar{\delta} \epsilon / \epsilon \left( x \right)$ is continuous and bounded in this interval, while the function $\epsilon \left( x \right)$ has a constant sign. Now, let us relate the variation $\bar{\delta} \epsilon $ to helioseismological data. The main source of energy inside the Sun is the proton-proton chain of thermonuclear reactions. The corresponding energy generation rate $\epsilon$ can be roughly estimated as $\epsilon \propto \rho X^2\, t^{\nu}\,$, where $\nu \approx 4.5,\; X \approx 1-Y$. Accounting for the equality (\ref{deltat}), one can write:
\begin{equation}\label{deps}
\frac{\bar{\delta} \epsilon}{\epsilon} = \frac{\bar{\delta} \rho}{\rho} + 9\, \frac{\bar{\delta} c_s}{c_s} + 4.5\,  \frac{\bar{\delta} \mu}{\mu} - 2\,  \frac{\bar{\delta} Y}{Y}.
\end{equation}
Mean molecular weight is related to helium abundance by the formula (\ref{3}), so $\bar{\delta} \mu / \mu = 5\, \bar{\delta} Y / \left( 8 - 5\, Y \right)$. Helioseismological data impose constraints on the variations of density and sound speed inside the core $\left| \bar{\delta} \rho / \rho \right| < 0.02,\; \left| \bar{\delta}  c_s / c_s \right| < 0.003$, as well as give the required variation of the initial helium abundance $\bar{\delta} Y \simeq 0.017$. Given that, one can infer $\left| \bar{\delta} \epsilon / \epsilon \right| < 0.07$. Then the equation (\ref{avev}) shows that $\left| \left( \bar{\delta} l / l \right)_{\text{nuc}} \right| < 0.07 \ll 1$, quod erat demonstrandum.

Relative variations of temperature, temperature derivative and density derivative in the case of the model with the opacity increase are given by the equalities analogous to the equalities (\ref{deltat}), (\ref{deltadt}), (\ref{deltadp}), respectively -- one just have to put bars over deltas, for we deal now with the second pair of models. In each pair of models there are the B16 Standard solar model and the solar model solving the solar abundance problem. The both model-solutions, be it a model with the increased opacity function or a model with the additional energy transfer, must reproduce the same helioseismological sound speed and density profiles, as well as the same helioseismological value for the surface abundance of helium. This means that  $\delta c_s = \bar{\delta} c_s$ and $\delta \rho = \bar{\delta} \rho $, which yields $\delta p' = \bar{\delta} p'$. Profiles of a variation of mean molecular weight in the two pairs of models are in general not similar, because a variation of abundance of helium is composed of two main contributors: a variation of the parameter of initial helium abundance $\delta Y_0$ and a variation of helium abundance due to the change in thermonuclear reactions rates $\delta Y_{\text{nuc}}$. From these two contributors, the surface abundance of helium is influenced only by the variation $\delta Y_0$. Moreover, the surface abundance is fixed by helioseismology, so the following equality should hold: $\delta Y_0 = \bar{\delta} Y_0$. The increased opacity model does not change the rate of the thermonuclear reactions, so $\bar{\delta} Y_{\text{nuc}} = 0$, while the model with additional energy transfer can change these rates in case $\bar{\epsilon} \neq 0$, that is why generally $\delta Y_{\text{nuc}} \neq 0$. Thus, relative variations of mean molecular weight in the two pairs of models considered are related in the following way:
\begin{equation}
\frac{\delta \mu}{\mu} = \frac{\bar{\delta} \mu}{\mu} + \left( \frac{\delta \mu}{\mu} \right)_{\text{nuc}}.
\end{equation}
Now, let us substitute all variations with bars in the right-hand side of the equality (\ref{beginop}) with the variations without bars according to the rules established above and subtract the equality that we get from the equality (\ref{deltal}). We obtain the following result:
\begin{equation}\label{subtr}
\frac{\delta l}{l} - \frac{\bar{\delta} \kappa}{\kappa} = 2 \beta \, \frac{\delta c_s}{c_s} - \alpha \, \frac{\delta \rho}{\rho} + \beta \, \frac{\bar{\delta} \mu}{\mu} + \left( 4+ \beta \right) \cdot \left( \frac{\delta \mu}{\mu} \right)_{\text{nuc}} + \frac{t}{t'} \left( \frac{\delta \mu}{\mu} \right)'_{\text{nuc}} + \left( \frac{\bar{\delta} l}{l} \right)_{\text{nuc}}.
\end{equation}

We note that the total change of luminous flux power in the model with additional energy transfer is composed from two parts: $\delta l = \left( \delta l \right)_{\text{nuc}} + \left( \delta l \right)_{\text{add}}\,$, where the first part originates from the variation of thermonuclear reactions rate inside the core due to the variations of temperature, density and mean molecular weight inside the core, while the second part originates from the additional energy transfer which we parameterized by the term $\sum_i \epsilon_i = \sum_i \tilde{\epsilon}_i + \bar{\epsilon}$ in the equation of production/loss of energy. The first part can be calculated given the parameter $\bar{\epsilon}$. We are interested in the additional energy transfer required to solve the solar abundance problem, so in the left-hand side of our equations we leave  solely the additional energy transfer profile $\left( \delta l \right)_{\text{add}}\,$. In case $\bar{\epsilon} = 0$, one gets $\delta c_s = \bar{\delta} c_s, \; \delta \rho = \bar{\delta} \rho, \; \delta Y = \bar{\delta} Y$, so, as it can be seen from the equations (\ref{avev}) and (\ref{deps}), $\left( \delta l \right)_{\text{nuc}} = \left( \bar{\delta} l \right)_{\text{nuc}}$.

Now, we write the equality (\ref{subtr}) in its final form:
\begin{equation}\label{deltalfull}
\begin{split}
\left( \delta l \right)_{\text{add}}\,= l \cdot \left( \frac{\bar{\delta} \kappa}{\kappa} + 2 \beta \, \frac{\delta c_s}{c_s} - \alpha \, \frac{\delta \rho}{\rho} + \beta \, \frac{\bar{\delta} \mu}{\mu} + \right. \qquad \qquad \qquad \qquad \qquad \qquad & \\ 
\left. \left( 4+ \beta \right) \cdot \left( \frac{\delta \mu}{\mu} \right)_{\text{nuc}}\! + 
\frac{t}{t'} \left( \frac{\delta \mu}{\mu} \right)'_{\text{nuc}} + \left( \frac{\bar{\delta} l - {\delta} l}{l} \right)_{\text{nuc}} \right) &, 
\end{split}
\end{equation}
which is valid at the point $m = 0$ as well. The last three terms of the equality (\ref{deltalfull}) are not zero only in case $\bar{\epsilon} \neq 0$ and only inside the core. 

\subsubsection{No net loss or gain of energy: $\bar{\epsilon} = 0$}\label{zero}

\begin{figure}[h!]
	\caption{Additional power of luminous flux, which allows one to solve the solar abundance problem, as a function of radial coordinate inside the core and the radiative zone, $\bar{\epsilon} = 0$}\label{dl1}
	\centering \includegraphics[height=8cm]{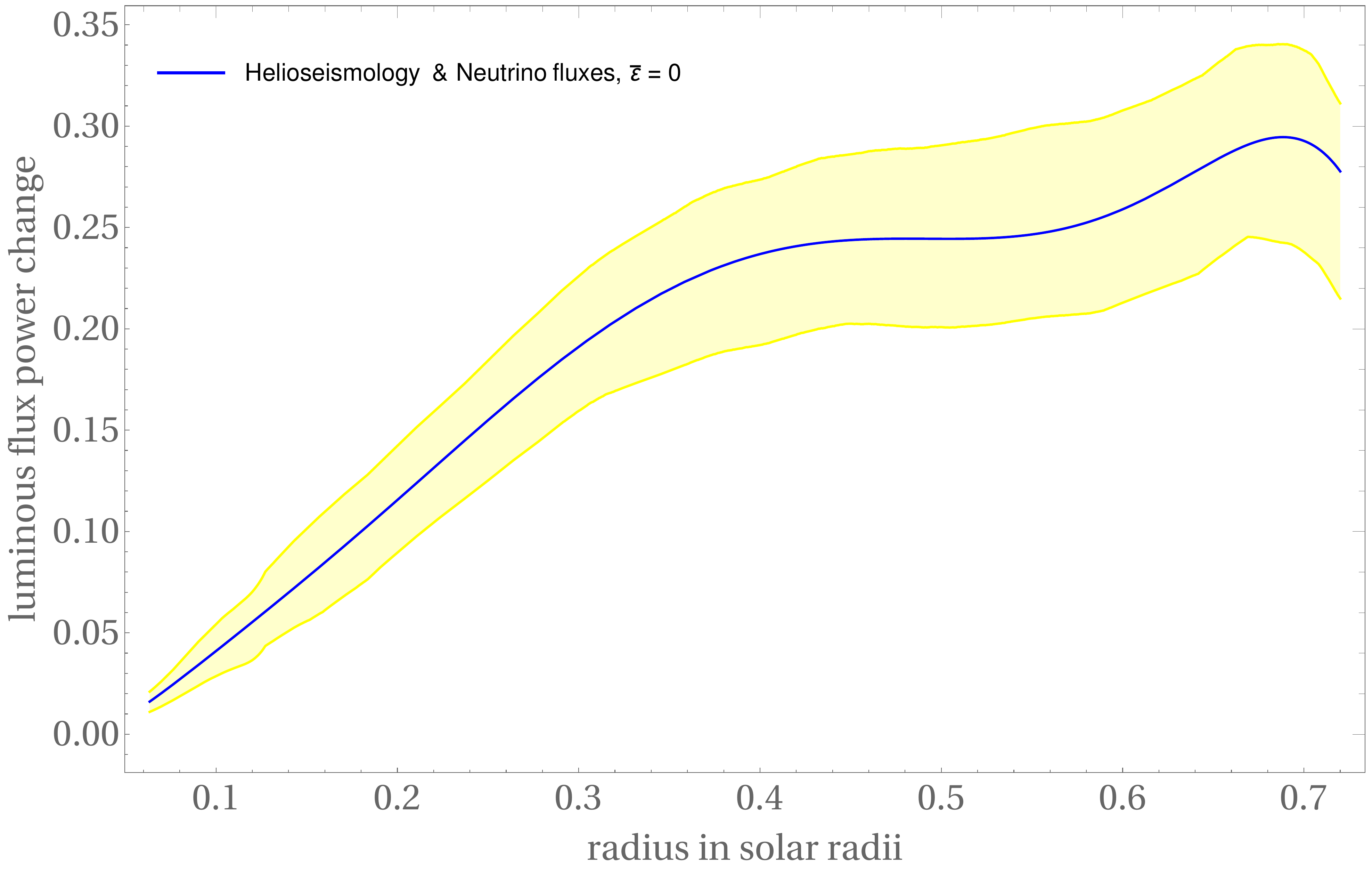}
\end{figure}

Let us first consider the case $\bar{\epsilon} = 0$ -- the Sun does not lose or gain any additional amount of energy:
\begin{equation}\label{deltal0}
\delta l = l \cdot \left( \frac{\bar{\delta} \kappa}{\kappa} + 2 \beta \, \frac{\delta c_s}{c_s} - \alpha \, \frac{\delta \rho}{\rho} + \beta \, \frac{\bar{\delta} \mu}{\mu} \right) .
\end{equation}
The quantity $\bar{\delta} \mu $ is fixed by the helioseismological surface helium abundance:
\begin{equation}\label{mumi}
\frac{\bar{\delta} \mu}{\mu} = \frac{5\, \delta Y_0}{8 - 5\, Y} = \frac{5\, \delta Y_{\text{s}}}{8 - 5\, Y}.
\end{equation}
We take the profile $\bar{\delta} \kappa$ from the work \cite{Villante:2015xla}, then we recalculate this profile as well as the profiles $\delta c_s/c_s$ and $\delta \rho / \rho$, given in the plots \ref{deltacs} and \ref{deltaro}, from radial coordinate to the mass one, using the dependence of radius on mass in the B16 Standard solar model. Such switch of coordinate is well justified, for a straightforward calculation of the integral (\ref{integ}) gives $\left( \delta r \right)_{\text{max}} = 0.003$, which means that the total error of $\delta l$ due to the coordinate change does not exceed $0.5\%$. Now that we have all the quantities from the right-hand side of the equality (\ref{deltal0}), we calculate the required variation of the profile of the luminous flux power $\left( \delta l \right)_{\text{add}}\,$. Let us find as well uncertainties of the calculated profile. First, as it was discussed in the section \ref{first}, uncertainties of the profiles $\delta \rho / \rho $ and $\delta c_s / c_s$ are both about several per mille. For there holds the inequality  $\alpha \ll 2\, \beta$ (see plots \ref{opal_a} and \ref{opal_b}), one can neglect uncertainties of the density profile. Uncertainties of the sound speed profile $\delta c_s / c_s$ are given in the work \cite{Vinyoles:2016djt}, errors of the opacity profile $\bar{\delta} \kappa / \kappa$ -- in the work \cite{Villante:2015xla}. Finally, uncertainties of the variation of the mean molecular weight profile $\bar{\delta} \mu / \mu $ can be calculated via the formula (\ref{mumi}) given the uncertainty of the theoretically calculated helium abundance in the B16 Standard solar model ($\text{err}\! \left( Y \right) \simeq 0.006$) and the uncertainty of helioseismological determination of the surface helium abundance $\text{err}\! \left( Y_{\text{s}}^{\text{hel}}\right)  \simeq 0.004$. Profile of the required variation of luminous flux power with its uncertainties is given in plot \ref{dl1} as a function of radial coordinate.

Let us note that the formula (\ref{deltalfull}) is valid only inside the core and the radiative zone, for we used the Fick's law which is not relevant in the convective zone. Nevertheless, it is easy to deduce that the convection zone does not have regions which emit or absorb significant amount of energy compared to the convective energy flows. The helioseismological sound speed profile in the convection zone agrees perfectly with the predictions of the Standard solar models, whereas the helioseismological helium abundance in the convection zone is higher than the calculations suggest. Due to the effective convective mixing, helium abundance in the convection zone is constant, therefore mean molecular weight is constant, too. Then the equality (\ref{deltat}) suggests that the temperature change must be constant as well. To the contrary, significant additional energy transfer inside the convection zone would lead to a change in the temperature gradient, i.e. to a non-uniform change of temperature. The contradiction we get shows that energy transfer inside the convection zone must be dominantly convective, with negligible contributions from the other means of energy transport. A boundary condition on the solar surface fixes the luminous flux power throughout the convection zone, thus $\delta l = 0$ there. In the context of a solution to the solar abundance problem via additional energy transfer, this means that the luminous flux power should fall near the boundary of the radiative and convection zones from the value of about $1.26\,L_{\odot}$ (see plot \ref{dl1}) to its surface value equal to the solar luminosity $L_{\odot}$. Thus, the solar model can be reconciled with helioseismological data if one assumes emission of some particles at the boundary between the radiative and convection zones, as well as heating of the inner part of the Sun within the approximate region $r \in [0.1,\, 0.3]$, see plot \ref{dl1}. 

\begin{figure}[h!]
	\caption{Additional luminous flux power which allows one to reconcile the theoretical sound speed profile with the helioseismological one as a function of radial coordinate inside the core and the radiative zone, $C = -0.26,\; \bar{\epsilon} = 0$}\label{dlDM1}
	\centering \includegraphics[height=8cm]{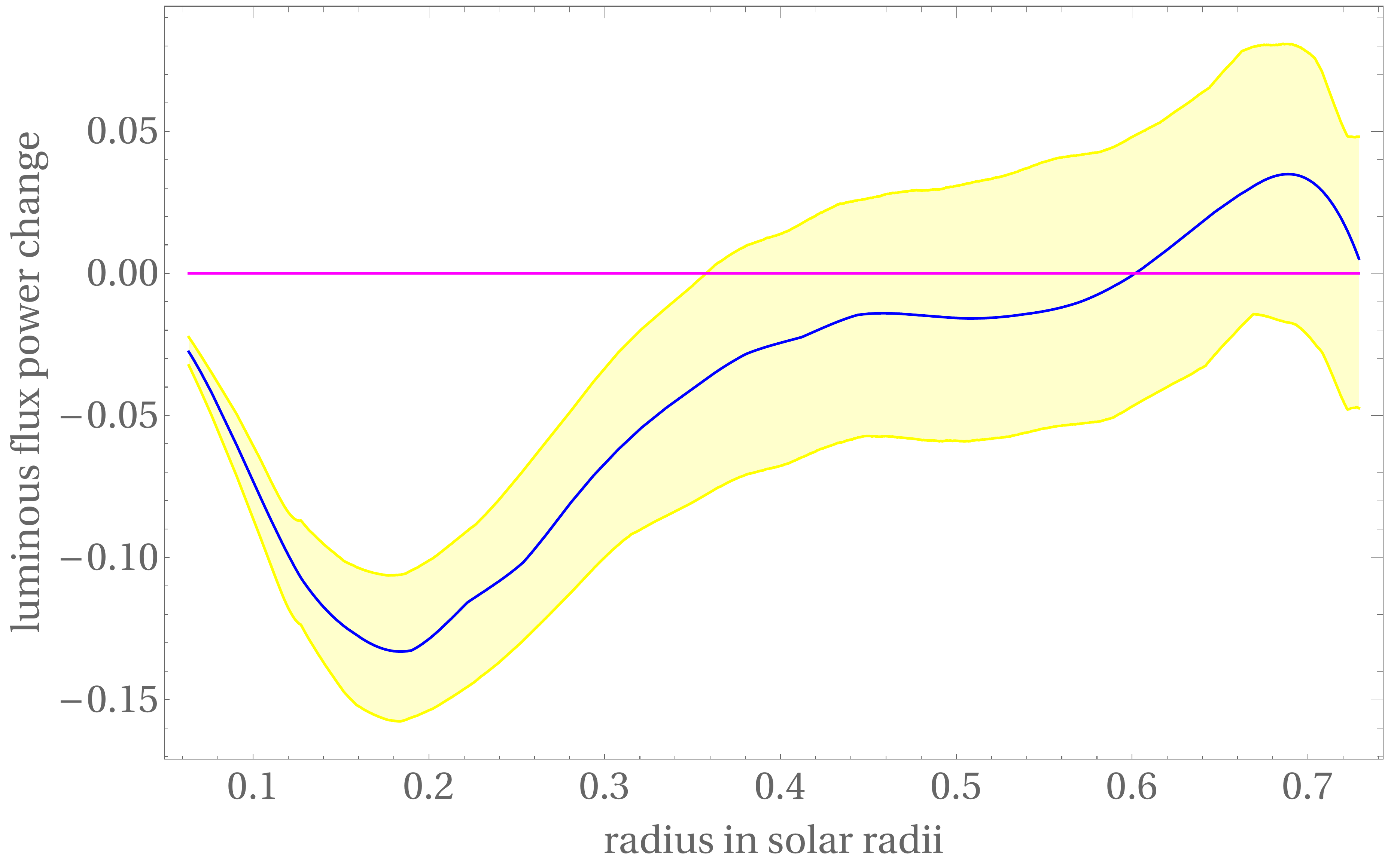}
\end{figure}

There is a connection of our work to studies of the effects which exhibit new hypothetical particles on the solar interior. As it was mentioned in the beginning of the section \ref{task}, global rescaling of the opacity does not influence sound speed profile, so if we are only interested in the reconciliation of the sound speed profile with helioseismology, we can add an arbitrary constant $C$ to the right-hand side of the equality (\ref{deltal0}). In case $C = -0.26$, the resulting power variation profile evidences additional energy transfer from the centre of the solar core to its periphery, see plot \ref{dlDM1}. This way to resolve the sound speed anomaly resembles the solution of this anomaly provided by dark matter diffusive energy transport inside the core, see \cite{Geytenbeek:2016nfg}.

\subsubsection{General case: $\bar{\epsilon} \neq 0$}

\begin{figure}[h!]
	\caption{Blue and yellow: required addition $\left( \delta l \right)_{\text{add}}$ to the power of luminous flux as a function of radial coordinate inside the core and radiative zone, $\bar{\epsilon} = - 0.2$; one can see that an additional loss of energy alone cannot be a solution to the solar abundance problem: $\left( \delta l \right)_{\text{add}} > 0$}\label{dlm1}
	\centering \includegraphics[height=8cm]{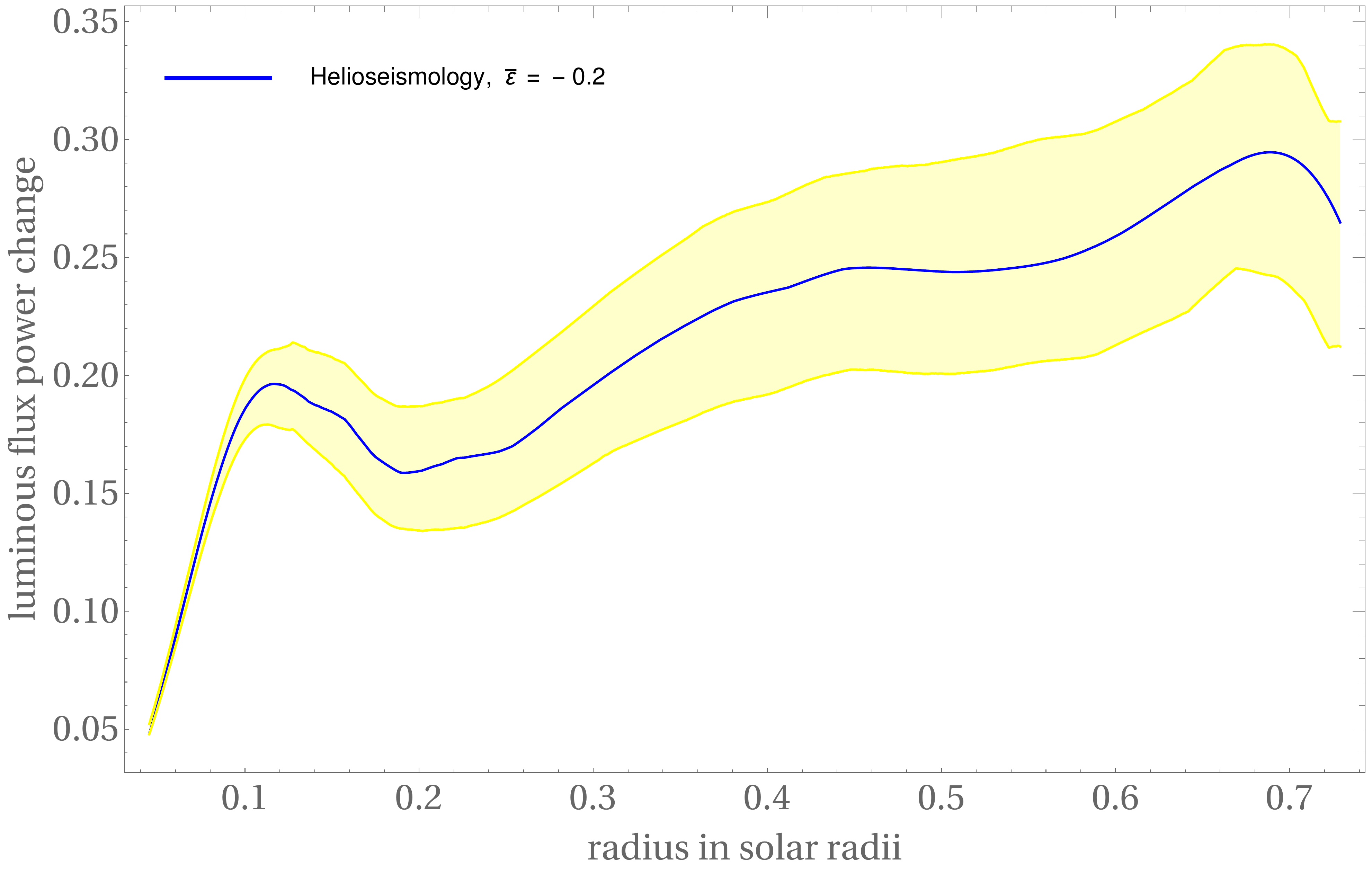}
\end{figure}

Now, let us consider the case $\bar{\epsilon} \neq 0$; for certainty, we assume $\bar{\epsilon} < 0$. In this case, the emission and absorption are not equilibrated, i.e. a part of the particles emitted leaves the Sun. This means that we have to account for the changes of the required power variation inside the core due to the last three terms in the equality (\ref{deltalfull}). These three additional terms can be estimated using the formulae (\ref{munuc}), (\ref{avev}), (\ref{deps}). In particular, we find:
\begin{equation}
\left( \frac{\delta l - \bar{\delta} l}{l} \right)_{\text{nuc}} = \frac{\xi_1\, \bar{\epsilon}}{l} \int\limits_0^m \frac{16-32.5\,Y}{8-5Y} \cdot \frac{Y_{\text{nuc}}}{Y} \cdot \epsilon \, dx \, < \, 3\cdot 10^{-3},
\end{equation}
so that this term is negligible and an error from the rough estimate (\ref{deps}) does not influence the final result. Required variation of luminous flux power in the case $\bar{\epsilon} = - 0.2$, corresponding to the maximum energy loss allowed by solar neutrino data, is given in the plot \ref{dlm1}.
We see that the additional power required is mostly positive. To the contrary, if the solar abundance problem could be solved by a loss of energy alone, the additional power profile would be non-positive. Thus, we conclude that emission of particles from the Sun, without any other modifications of solar physics, cannot solve the solar abundance problem. The case $\bar{\epsilon} > 0$, the energy hypothetically being absorbed inside the Sun from outside, can be analyzed analogously to the case $\bar{\epsilon} < 0$. Again, the solar abundance anomaly cannot be explained by deposition of energy alone, otherwise the required addition to the flux power $\left( \delta l \right) _{\text{add}}> 0.2$ in the upper layers of the radiative zone would contradict the solar neutrino data which constrain the net gain of energy by the Sun $l_{\text{gain}} < 0.1$. Therefore, additional energy transfer solutions to the solar abundance problem have to include both new sources and new sinks of energy. The net power gained or lost cannot exceed $0.1\,L_{\odot}$; thus the main mechanism of additional energy transfer has to be similar to the one described in the previous subsection~\ref{zero}. Although some limited loss or gain of energy by the Sun is in principle allowed, the required addition to the power profile becomes quite wavy inside the core (see Fig. \ref{dlm1}), which can be difficult to explain within a real physical model. 

\section{Particle physics model}

\subsection{Resonant hidden photon emission}\label{hpemi}

In order to solve the solar abundance problem in the context of the additional energy transfer discussed in the section \ref{second} of this work, we need significant emission of energy from the region of the boundary of the solar radiative zone. Such localized emission can emerge from resonant processes due to the mixing of WISP particles (weakly interacting slim particles, see \cite{Jaeckel:2010ni}) with photons. WISPs include such hypothetical new particles as axions, axion-like particles, hidden photons, chameleons etc. The existence of these particles is motivated from the theoretical considerations, coming from the string theory and/or from more direct attempts at the solution of some particle physics problems. Various laboratory experiments and astrophysical observations showed that if these particles exist in nature, their non-resonant interaction with ordinary matter is extremely weak. Thus, the only significant interaction between these particles and solar plasma can happen within the resonant regime. In particular, in case of hidden photons, the resonant conversion occurs whenever the thermal photon mass $\omega_{pl}$ is equal to the hidden photon mass $m_{\gamma'}$ or to the hidden photon energy $\omega_{\gamma'}$, depending on the polarization of the hidden photon (transverse or longitudinal, respectively); in this work we set the light speed equal to unity.

Many extensions of the Standard model naturally incorporate an additional $U\! \left( 1 \right)'$ gauge symmetry, besides the electromagnetic one. If there are several $U\! \left( 1 \right)$ gauge groups in the model, the corresponding gauge fields $A_{1\mu}$ and $A_{2\mu}$ mix:
\begin{equation}\label{l0}
L_0 = -\frac{1}{4}F_{1\mu \nu} F_1^{\mu \nu} - \frac{1}{4}F_{2\mu \nu} F_2^{\mu \nu} - \frac{\chi}{2}F_{1\mu \nu} F_2^{\mu \nu} .
\end{equation}
After the diagonalization of the Lagrangian one can separate the fields corresponding to the two particles, which are ordinary and hidden photons. The mass of the ordinary photon must be equal to zero, which is not necessarily true for the mass of the hidden photon. One can give mass to the hidden photon e.g. by the Stueckelberg mechanism \cite{Stueckelberg:1900zz, osti_4006195}:
\begin{equation}\label{lmass}
\begin{split}
L_{\text{Mass}} = \frac{1}{2} \left( \partial_{\mu} \sigma  \right. & \left. + M_1 A_{1\mu} + M_2 A_{2\mu} \right)^2 = \\
& \frac{1}{2} M_1^2 A_{1 \mu} A_1^{\mu} +\frac{1}{2} M_2^2 A_{2\mu} A_2^{\mu} + M_1 M_2 A_{1 \mu} A_2^{\mu},
\end{split}
\end{equation}
where the gauge is chosen where $\sigma = 0$. After the diagonalization of the Lagrangian (\ref{lmass}) one of the fields becomes massless, while another gets the mass $m_{\gamma'} = \sqrt{M_1^2 + M_2^2}\,$. If $m_{\gamma'} \neq 0$, the ordinary photon can convert into the hidden one due to the kinetic mixing. Further, we consider the special case of the model outlined above where the mass mixing parameter $\epsilon \equiv M_2/M_1$ is negligible compared to the kinetic mixing $\chi$: $\epsilon \ll \chi \ll 1$. In the absence of mixing in the  mass sector $M_2 = 0$. The production of hidden photons inside solar plasma in this case, including a resonant as well as a non-resonant regime, was discussed in the works \cite{An:2013yfc, Redondo:2008aa}. In the mass region where the resonant production is possible, it surpasses by far the non-resonant one. The condition for the resonant production of longitudinally polarized hidden photons is $\omega_{\gamma'} = \omega_{pl}$, so at any point of the Sun where plasma frequency satisfies the inequality $\omega_{pl} \geq m_{\gamma'}$ there is production of the hidden photons with some definite energy. Such resonant production encompasses a large region inside the Sun. On the contrary, the resonant production of transversely polarized hidden photons is characterized by the resonant region which is a very thin spherical shell, while the maximum of the resonance is reached at the same sphere for any energy of the hidden photons produced. Let us estimate the size of such a resonant shell if the resonance happens close to the radiative zone boundary:
\begin{equation}\label{width}
\Delta r_{res} = \left( \Delta \omega_{pl}^2 \right)_{res} \cdot \left( \frac{d \omega_{pl}^2}{dr} \right)^{-1} = 2\, \omega_{\gamma'}\, \kappa \rho \cdot \frac{\kappa_0\, m_e\, m_H}{2\pi \alpha \left( X+1 \right)} \left| \frac{d\rho}{dr} \right|^{-1} \sim 10^{-4}, 
\end{equation}
for the width of the resonance is $\left( \Delta \omega_{pl}^2 \right)_{res} = 2\, \omega_{\gamma'}\, \Gamma = 2\, \omega_{\gamma'}\,  \kappa \rho \cdot \kappa_0 \rho_0 $ (see \cite{Redondo:2008aa}), plasma frequency squared is $ \omega_{pl}^2 = \frac{2\pi \alpha (X+1)}{m_e m_p} \rho \rho_0$, where we used the fact that the number density of electrons inside solar plasma is related to the one of protons by the expression $n_e \simeq n_H \left( 1+ X \right) / 2X$, $\alpha$ is the fine structure constant. We see that the width of the resonant shell is negligible compared to the solar scale, so one can consider it a sphere. Such narrow resonant emission of energy close to the radiative zone boundary $r \simeq 0.7$ is exactly what we need in order to implement the first part of the additional energy transfer solution to the solar abundance problem, discussed in the section \ref{second}. The mass of the hidden photon corresponding to the emission from this region is $m_{\gamma'} = \omega_{pl} \left( r=0.7 \right) = 12\,\text{eV}$. As it can be inferred from the work \cite{Redondo:2013lna}, in particular from the Fig. 3 there, the resonant production of longitudinally polarized hidden photons is negligible compared to the resonant production of transversely polarized ones in this scenario, so one can consider the resonant sphere discussed above as the only source of hidden photons inside the Sun: all other possible sources are too faint in case $m_{\gamma'} \simeq 12\, \text{eV}$. The power emitted can be calculated following the formula derived in the work \cite{Redondo:2008aa}: $L_{\gamma'} \propto r^2 t^3 m_{\gamma'}^4 \cdot \left( d \omega_{pl}^2 / dr \right) ^{-1}$. In our normalized quantities the power emitted is given by the following expression:
\begin{equation}
l_{\gamma'} = \frac{\chi ^2 M_{\odot} T_0^3}{m_e m_H L_{\odot}} \cdot \frac{24 \zeta \! \left( 3 \right)}{\pi} \, \alpha \left( X+1 \right) \rho^2 r^2 t^3 \left| \frac{d\rho}{dr} \right|^{-1}\! = 0.26 \cdot \left( \frac{\chi}{1.5\cdot 10^{-13} } \right) ^2 .
\end{equation}
Thus, if $\chi \simeq 1.5\cdot 10^{-13}$ we get the required value $\Delta \left( \delta l \right) \simeq -0.26$ for the sharp decrease of luminous flux near the radiative zone boundary, see plot \ref{dl1} and remember that there must hold $\delta l = 0$ in the convection zone. 

It is remarkable that the indicated value of the kinetic mixing $\chi \simeq 1.5\cdot 10^{-13}$ given the hidden photon mass $m_{\gamma'}=12\, \text{eV}$ does not conflict with any of the existing laboratory and astrophysical constraints, apart from those coming from the Sun, as it can be clearly seen from the exclusion plot in the Fig. 3 of the work \cite{Redondo:2013lna}. The constraints coming from the Sun are derived from the energy loss argument under the assumption $l_{\gamma'} = l_{loss}$: in order not to contradict the values of the observed neutrino fluxes one has to impose $l_{\text{loss}}<0.1$, as we mentioned in the section \ref{second}. Thus, these constraints are no longer valid if $l_{\gamma'} > l_{loss}$, i.e. if some of the energy emitted is compensated by additional heating of the Sun, therefore providing for additional energy transfer, while losses of energy are small enough. The results of the section \ref{second} show that in case one considers the solar abundance problem as an indication of the additional energy transfer or loss, the transfer must be indeed quite large while the losses must be small, thus justifying the chosen value for the kinetic mixing  $\chi \simeq 1.5\cdot 10^{-13}$.

\subsection{Millicharged particles}\label{mil}

Let us add to the Lagrangian (\ref{l0}), (\ref{lmass}) considered in the previous section an interaction of the gauge fields with the external currents $J_{\mu}$ and $J_{\mu}'$:
\begin{equation}\label{Js}
L_{\text{int}} = J_{\mu} A_2^{\mu} + J_{\mu}' A_1^{\mu},
\end{equation}
where the first current is responsible for the interaction with the Standard model particles, while the second -- for the interaction with the particles from the hidden sector. We still assume the kinetic mixing prevails over the mass one, so that we set $M_2 \equiv 0$ in our calculations. We diagonalize the kinetic sector of the model via tranformations $A_{\gamma}^{\mu} = A_2^{\mu}$, $A_{\gamma'}^{\mu} = A_1^{\mu} + \chi A_2^{\mu}$, where $A_{\gamma'}^{\mu}$ is a hidden photon field, $ A_{\gamma}^{\mu}$ is an ordinary photon field. The interaction and mass parts of the Lagrangian get the following form:
\begin{equation}\label{interaction}
\begin{split}
L_{\text{int}} &= J_{\mu}' A_{\gamma'}^{\mu}\, + \left( J_{\mu} - \chi J_{\mu}' \right) A_{\gamma}^{\mu} , \\
L_{\text{Mass}}\, =&\:\: \frac{1}{2} m_{\gamma'}^2 A_{\gamma' \mu} A_{\gamma'}^{\mu} - \chi m_{\gamma'}^2 A_{\gamma' \mu} A_{\gamma}^{\mu}.
\end{split}
\end{equation}
In case energy scale is larger than the hidden photon mass $m_{\gamma'} = 12\, \text{eV}$, the hidden sector particles entering the current $J_{\mu}'$ acquire an effective millicharge $\varepsilon = \chi e' / e$, where $e'$ is the gauge coupling constant of the hidden sector. This millicharge must be very small not to contradict various experiments and observations, see \cite{Davidson:2000hf}. Let us suppose that twice the mass $m_c$ of the millicharged particles is less than the hidden photon mass: $ 2\, m_c < m_{\gamma'} = 12\, \text{eV}$. In this case hidden photons are no longer stable and decay into millicharges. We estimate this decay length in the laboratory frame of reference:
\begin{equation}\label{d}
d \simeq \frac{\left\langle \gamma_{\gamma'} \right\rangle}{\Gamma_{\gamma'}} = \frac{2 \left\langle \omega_{\gamma'} \right\rangle}{\alpha' m_{\gamma'}^2} = \frac{3\, t\, T_0}{\alpha' m_{\gamma'}^2} = 0.01\, R_{\odot} \cdot \left( \frac{e'}{10^{-6}} \right)^{-2},
\end{equation}
where $\alpha' \equiv e'^2/4\,\pi$; $\,\gamma_{\gamma'}$ is a Lorentz-factor of the hidden photon, factor two in the numerator corresponds to the two transverse polarizations of the hidden photon; we assumed $m_c \ll m_{\gamma'}$ for this estimate. It may seem confusing at the first sight, but the average energy of the hidden photons is $\left\langle  \omega_{\gamma'} \right\rangle \simeq 1.5\,T$, which is by a factor of two less than the average energy of the ordinary photons. The spectrum of the hidden photons is shifted to lower energies, because the probability of the resonant conversion decreases as a function of energy, see \cite{Redondo:2008aa}. Indeed, following that work, one can write the distribution of the transverse resonant hidden photons with energy:
\begin{equation}\label{distrib}
\begin{split}
\frac{dN_{\gamma'}}{d\omega} = 8\pi  \chi^2 m_c^4 r^2 \cdot \frac{\sqrt{\omega^2 - m_{\gamma'}^2}}{e^{\omega / T} - 1} \cdot & \left| \frac{d\omega_{pl}^2}{dr} \right|^{-1} \cdot R_{\odot}^3 = \\
& \frac{4\, \chi^2 R_{\odot}^3 m_e m_H m_c^4}{\alpha \left( X+1 \right) \rho_0 } \cdot \frac{\sqrt{\omega^2 - m_{\gamma'}^2}}{e^{\omega / T} - 1}  \cdot r^2 \left| \frac{d\rho}{dr} \right|^{-1} ,
\end{split}
\end{equation}
which allows us to easily calculate the average energy $\left\langle  \omega_{\gamma'} \right\rangle$:
\begin{equation}
\left\langle  \omega_{\gamma'} \right\rangle = \frac{\int_{m_{\gamma'}}^{\infty} d\omega \, \omega \, \frac{dN_{\gamma'}}{d\omega}}{\int_{m_{\gamma'}}^{\infty} d\omega \, \frac{dN_{\gamma'}}{d\omega}} = \frac{12\zeta \! \left( 3 \right)}{\pi^2}\, T \simeq 1.5\, T.
\end{equation}
We neglected the small quantity $m_{\gamma'}^2/T^2 \ll 1$. From the estimate (\ref{d}) one can see that in case $e' \gtrsim 10^{-6}$ the distance travelled by a hidden photon before its decay is negligibly small compared to the solar scale. The latter allows us to consider the region of the production of millicharges a sphere with a great accuracy. This sphere is the only significant source of millicharges inside the Sun, for the dominant contribution to the production of millicharges is provided by on-shell hidden photons while practically the only source of the hidden photons is a thin resonant shell discussed in the section \ref{hpemi}. Direct production of millicharges through ordinary photons ($\sim \varepsilon^2 \alpha^2$) is suppressed due to the smallness of millicharge $\varepsilon$ and fine-structure constant $\alpha$, see section \ref{heating} for the discussion of electromagnetic interactions of millicharges with the solar plasma.

\subsection{Interaction of millicharges with solar magnetic fields}\label{magnetic}
Besides high temperatures which provide us with a large number of photons inside the solar plasma and thus with effects of the hypothetical hidden $U\left( 1 \right)'$ symmetry, the solar plasma has another feature -- large magnetic fields. The millicharged particles of the model under consideration interact with the solar magnetic fields, so let us discuss what it is known about the structure of the magnetic field inside the core and the radiative zone of the Sun. Some information about it was extracted from the helioseismological data. We have not yet mentioned that apart from the parameters used by us to determine the required luminous flux power profile, the solar oscillation spectrum allowed one to infer the profile of the differential rotation of the Sun \cite{1988ESASP.286..149C}. It was found that while the rotation of different layers inside the solar convective zone is highly non-uniform, the radiative zone of the Sun rotates as a rigid body. The transition layer between these two regions of the different rotation regimes is very thin -- the estimates of its width are $\Delta = \left( 0.02 - 0.05 \right) R_{\odot}$ \cite{tachocline}. The described rotation profile largely contributes to our knowledge about the inner solar magnetic field due to the Ferraro isorotation law \cite{1937MNRAS..97..458F}, which states that in a steady state of plasma the angular velocity is constant along the magnetic field lines. In particular, if there had been any magnetic field lines which cross the boundary of the radiative zone, then the differential rotation of the convection zone would be surely transmitted to the radiative zone during the lifetime of the Sun \cite{1999ApJ...519..911M}. Thus, the existence of a sharp transition from one rotation regime to another suggests the phenomenon of the magnetic field confinement: the global-scale interior magnetic field of the radiative zone must not penetrate into the convection zone \cite{1998Natur.394..755G}. This interior magnetic field is a fossil field that is no way related to the magnetic fields of the solar convection zone which are responsible for the observed solar magnetism phenomena and which are being constantly produced and amplified by the solar dynamo mechanism in the upper part of the narrow tachocline region \cite{1997ApJ...486..502C,2001ApJ...551..536D}. Due to the phenomenon of the confinement the magnetic field of the radiative zone must have a toroidal structure: the poloidal component must be largely suppressed \cite{1996ApJ...458..832B}. Very little is known about the strength of the toroidal field, the only upper bounds coming from the implications of the pressure this field would exhibit on the solar plasma, see \cite{1996ApJ...458..832B}. These upper bounds range from several MG near the radiative zone boundary \cite{2003ESASP.517...71R} till several tenth of MG in the inner part of the radiative zone \cite{2003ApJ...599.1434C}. We note that if there is a pressure of yet unknown origin exhibited on the magnetic fields within the radiative zone these bounds can be relaxed.

Let us estimate the Larmor radius of a millicharged particle with the energy $\omega_c$ inside the interior magnetic field of the Sun:
\begin{equation}\label{rl}
r_L = \frac{\sqrt{\omega_{c}^2 - m_c^2}}{\chi e' B} = 0.01\, R_{\odot}\cdot \left( \frac{\omega_c}{\text{keV}} \right) \left( \frac{B}{0.7\, \text{MG}} \right)^{-1} \left( \frac{\varepsilon}{7\cdot 10^{-15}} \right)^{-1}.
\end{equation}
We neglected the mass of the millicharges $m_c$, because $m_c < m_{\gamma'} \ll T$, and expressed the product $\chi e'$ in terms of millicharge $\varepsilon = \chi e'/e$. The best existing constraints on millicharge are the ones derived from astrophysics $\varepsilon \lesssim 10^{-14} - 10^{-13}\,$ \cite{Vinyoles:2015khy, Davidson:2000hf}. The energy of the millicharged particles which are produced near the radiative zone boundary is half the energy of the corresponding hidden photons $\left\langle \omega_{c} \right\rangle = \left\langle \omega_{\gamma'} \right\rangle /2 \simeq 0.75\, T \simeq 150\, \text{eV}$. Thus, one can clearly see from the formula (\ref{rl}) that for the allowed values of millicharge $\varepsilon \sim 10^{-15} - 10^{-14}$ the Larmor radius of the millicharged particles is very small compared to the solar structure scales. This means that produced millicharges are captured by the magnetic field of the radiative zone and start to drift along the magnetic field lines. As it was discussed in the previous paragraph, these field lines do not enter the convection zone, so that the millicharged particles are trapped within the solar interior. The millicharged particles accumulate in the regions of the higher field, which should be situated closer to the solar centre  \cite{1996ApJ...458..832B}, not in the poles, for the dominant component of the field must be a toroidal one. We have to note that the synchrotron radiation of the millicharged particles is completely negligible being proportional to the tremendously small number $\left(\varepsilon e\right) ^4$ or $\left(\varepsilon e\right) ^{8/3}$ in the classical and quantum regime, respectively \cite{1954Sokolov, Galtsov:2015rcs}; the quantum regime being realized for the masses of the millicharged particles $m_c \ll 10^{-3}~\text{eV}$ (assuming $B = 1~\text{MG}$, $\varepsilon = 7\cdot 10^{-15}$, $\omega_{c} = 150~\text{eV}$). In comparison, Coulomb interaction of the millicharged particles with the solar plasma that we consider below in the section \ref{heating} is of order $\varepsilon^2 e^4$.

\subsection{Solar plasma heating}\label{heating}

In the previous sections we have showed that the hidden photon model can provide one with the abrupt change in the luminous flux near the radiative zone boundary as well as with the emergence of millicharged particles which can be trapped after their production by the toroidal magnetic fields of the solar radiative zone. In order that this model could represent a solution to the solar abundance problem, one needs a mechanism which transfers energy from millicharges to the solar plasma. Moreover, the energy deposition should match the energy loss in hidden photons which we discussed in the section \ref{hpemi}, so that the net gain or loss of energy by the Sun is small. Suppose that the millicharged particles settle into an equilibrium configuration inside the solar core and transfer heat to the solar plasma. For this mechanism to succeed, the temperature $T_c$ of the gas of the dark sector particles has to be larger than the temperature of the solar plasma in the region of the supposed energy transfer, so that the heat is transferred in a right direction. This condition implies $T_c \gtrsim 1\, \text{keV}$. The average energy of the millicharged particles produced near the radiative zone boundary is $\left\langle \omega_{c} \right\rangle \simeq 150\, \text{eV}$, that is why an additional source of energy is needed which could maintain the temperature of the millicharged particles high enough. While we do not study such a mechanism in this work in detail, it seems to us that one possibility is to invoke some stable heavy dark particles charged under the hidden $U(1)'$ group and to suppose that these particles are gravitationally clumped within the centre of the Sun, radiating via dark photons. Examples of a similar construction include an asymmetric dark star \cite{Kouvaris:2015rea} or a mirror star \cite{Foot:1999hm, Curtin:2019lhm} at a sufficiently high temperature. After its emission, the dark radiation converts soon into millicharges, which cannot freely escape due to the interaction with the magnetic fields of the solar core. An envelope which contains the millicharged gas forms, where the gas pressure is equilibrated with the magnetic field pressure. The hidden photon luminosity of the central source can be crudely estimated as $ L_{\gamma'} \sim \sigma \, T_c^4 \,S_{env} \sim \pi^2 / 60 \cdot T_c^4 \cdot 4\pi \left( 0.4\, R_{\odot} \right)^2$; then the corresponding photon luminosity is  $ L_{\gamma'} \cdot P_{\gamma' \rightarrow \gamma} \sim \chi^2 m_{\gamma'}^4/\omega_{pl}^4 \cdot L_{\gamma'} \sim 10^{-17}\cdot L_{\odot}\; \ll \; L_{\odot}$, where we used $T_c \sim \text{keV}$; $\sigma $ is the Stefan-Boltzmann constant. Even given the crudeness of our estimates, one can see that  the photon production is absolutely negligible, so that there is no additional energy flow to the solar plasma associated with the $\gamma' \rightarrow \gamma$ conversion. Then let us suppose that the magnetic field strength distribution in the solar core is approximately spherically symmetrical and that the millicharged particles are fermions. The equilibrium configuration of the millicharges is then locally characterised by a Fermi-Dirac distribution with the temperature $T_c\left( r \right)$, which under the assumption of spherical symmetry depends only on the radial coordinate. 

Now let us consider the interactions between the millicharged particles and the solar plasma. The parameters of the Lagrangian \ref{interaction} that are favoured by the considerations of the previous sections are $\chi \sim 10^{-13}$ for kinetic mixing, $\varepsilon \sim 10^{-15} - 10^{-14}$ for millicharge. Then it follows that the dominant interaction between millicharges and solar plasma is Coulomb scattering ($\sim \varepsilon^2 \alpha^2$), the other interactions being suppressed by additional powers of $\varepsilon^2$ or $\alpha$. One has to note that charged particles propagating through plasma can yield Vavilov-Cerenkov emission of longitudinal plasmons \cite{PhysRev.123.711}, in case the speed of the particles is greater than the thermal speed, as well as transition radiation \cite{Ginzburg:1945zz}, in case the density of plasma changes along the trajectory of a particle. However, in our case the chemical potential of the millicharge gas is zero $n_c = n_{\bar{c}}$, so there are no millicharge currents and neither Vavilov-Cerenkov nor transition radiation.

The main process contributing to the energy transfer from millicharged particles to solar plasma is the process of Coulomb scattering of millicharges on electrons. The scattering on more heavy plasma particles is not significant, for the energy transfer is inversely proportional to the mass of the particle. Due to the smallness of the masses of the millicharged particles, the kinematics of the scattering coincides with the Compton scattering kinematics. The energy transfer in one act of scattering can be deduced from the 4-momentum conservation $p_e + p_c - k_c = k_e$, where $p$ are incoming momenta, $k$ -- outgoing. We square this identity to get:
\begin{equation}\label{start}
E_e \left( \omega_c - \omega_c' \right) - \left| \vec{p}_e \right| \left( \omega_c \cos \theta_1 - \omega_c' \cos \theta_2 \right) - \omega_c \, \omega_c' \left( 1 - \cos \theta \right) = 0,
\end{equation}
where $\omega_c$ and $\omega_c'$ are the energies of incoming and outgoing millicharged particles, respectively; $\theta_1$ and $\theta_2$ are the angles between an incoming electron and an incoming and outgoing millicharged particle, respectively; $\theta$ is the scattering angle. Taking into account that the electrons are non-relativistic and that the main contribution comes from the small values of $\theta$, one can write the following simplifying expressions: $E_e = m_e$, $\; \left| \vec{p}_e \right| = \sqrt{3\, m_e T}$, $\; \theta_1 - \theta_2 \simeq \theta \ll 1$, $\; \left| \omega_c' - \omega_c \right| \ll \omega_c$. Now, having simplified the equality (\ref{start}), let us find the change of energy of the millicharged particle:
\begin{equation}
\Delta \omega_c = \omega_c\, \sqrt{\frac{3\,T}{m_e}} \cdot 2 \sin \theta_1 \sin \frac{\theta}{2} - \frac{\omega_c^2}{m_e} \left( 1 - \cos\theta \right).
\end{equation}
The first term in the right-hand side of this equality does not have a definite sign, for $\sin\theta_1 \in [-1, 1]$. The angle $\theta_1$ in each collision is uniformly distributed, so the energy transfer process can be considered a random walk. The number of the scatterings required to heat millicharge or plasma due to the first term is given by $N \propto T \sin^2(\theta/2) / m_e $. Then the energy change can be written as:
\begin{equation}\label{gain}
\Delta \omega_c = \frac{\omega_c}{m_e} \cdot \left( \xi T - \omega_c \right) \cdot \left( 1 - \cos\theta \right),
\end{equation}
where $\xi$ is some constant that we will determine later.

The differential cross-section of the process under consideration has the form of the Mott cross-section of the Coulomb scattering of a light relativistic particle on a heavy particle at rest (the velocity and the recoil of the electron can be neglected):
\begin{equation}\label{mott}
d\sigma = \frac{\varepsilon^2 e^4}{8\pi \omega_c^2} \cdot \frac{\left( 1-x/2 \right) dx}{x^2},
\end{equation}
where $x = 1 - \cos\theta$; hereinafter, we use rationalized electromagnetic units. We have to integrate the product of (\ref{gain}) and (\ref{mott}) over the angle:
\begin{equation}
\int \Delta \omega_c \, d\sigma \propto \int\limits_0^2 \frac{1-x/2}{x + \psi} \, dx = \left( 1+ \frac{\psi}{2} \right) \cdot \ln \frac{2+\psi}{\psi} - 1,
\end{equation}
we have regularized the integral by adding a small parameter $\psi$ to the denominator. This regularization accounts for the non-zero mass of a mediator, which plays a significant role in case $x \ll 1$, i.e. small scattering angles. The effective non-zero mass of a photon arises due to the presence of screening in the medium. Due to smallness of the millicharge $\varepsilon$ and relativistic nature of incoming particles, the classical scattering theory does not apply and the minimum scattering angle is determined by diffraction effects: $\theta_{min} = \lambda_B/\lambda_{scr}$, where $\lambda_B = 1/ \omega_c$ is de Broglie wavelength of the particle, $\lambda_{scr} = 1/k_D$ is size of the screened region. For relativistic particles, the screening parameter $k_D$ equals plasma frequency, so finally we get the following expression for the regularization parameter: 
\begin{equation}
\psi = x_{min} = 1 - \cos\theta_{min} = \lambda_B^2/\left( 2\, \lambda_{scr}^2 \right) = \omega_{pl}^2/\left( 2 \omega_c^2 \right) .
\end{equation}
Let us note that everywhere inside the Sun $\psi \ll 1$. The value for the regularization parameter can be also obtained by calculating the cross-section (\ref{mott}) with the account for the mediator mass $m_{\gamma} = \omega_{pl}$ from the beginning. The propagator gives us the multiplier $1/\left( t_m -\omega_{pl} ^2 \right)^2$. When we substitute into this multiplier the Mandelstam variable $t_m = -2\, \omega_c^2 x$, we get the term in the denominator corresponding to $\psi$: 
\begin{equation}
\frac{1}{\left( t_m - \omega_{pl}^2 \right)^2 } \propto \frac{1}{\left( x + \omega_{pl}^2/\left(2\omega_c^2 \right) \right)^2 } \quad \Rightarrow \quad \psi = \frac{\omega_{pl}^2}{2\, \omega_c^2}.
\end{equation}
 We discuss the regularization parameter $\psi$ in such a detail, for some authors obtained different results. In particular, in the work \cite{Davidson:2000hf}, where they consider inter alia the interaction of the relativistic millicharged particles with matter in the context of the SN 1987A explosion, the screening scale for the calculation of the transport cross-section was chosen in a different way: $k_D^2 = \omega_{pl}^2/v^2 = 4\,\pi \alpha n_p / T$, which is indeed a screening scale for the particles moving with the non-relativistic speeds comparable to or less than the speeds of plasma protons $v \sim \sqrt{T/m_p}\, \ll 1$ . However, a particle moving through such a plasma with the speed of light should experience less screening: $k_D^2 = \omega_{pl}^2$, for the screening scale is inversely proportional to the relative speed.

\begin{figure}[h!]
	\caption{$ \text{K} \left( a \right) $ as a function of radial coordinate for the three values of temperature of the millicharge gas $T_c$}\label{K}
	\centering \includegraphics[height=8cm]{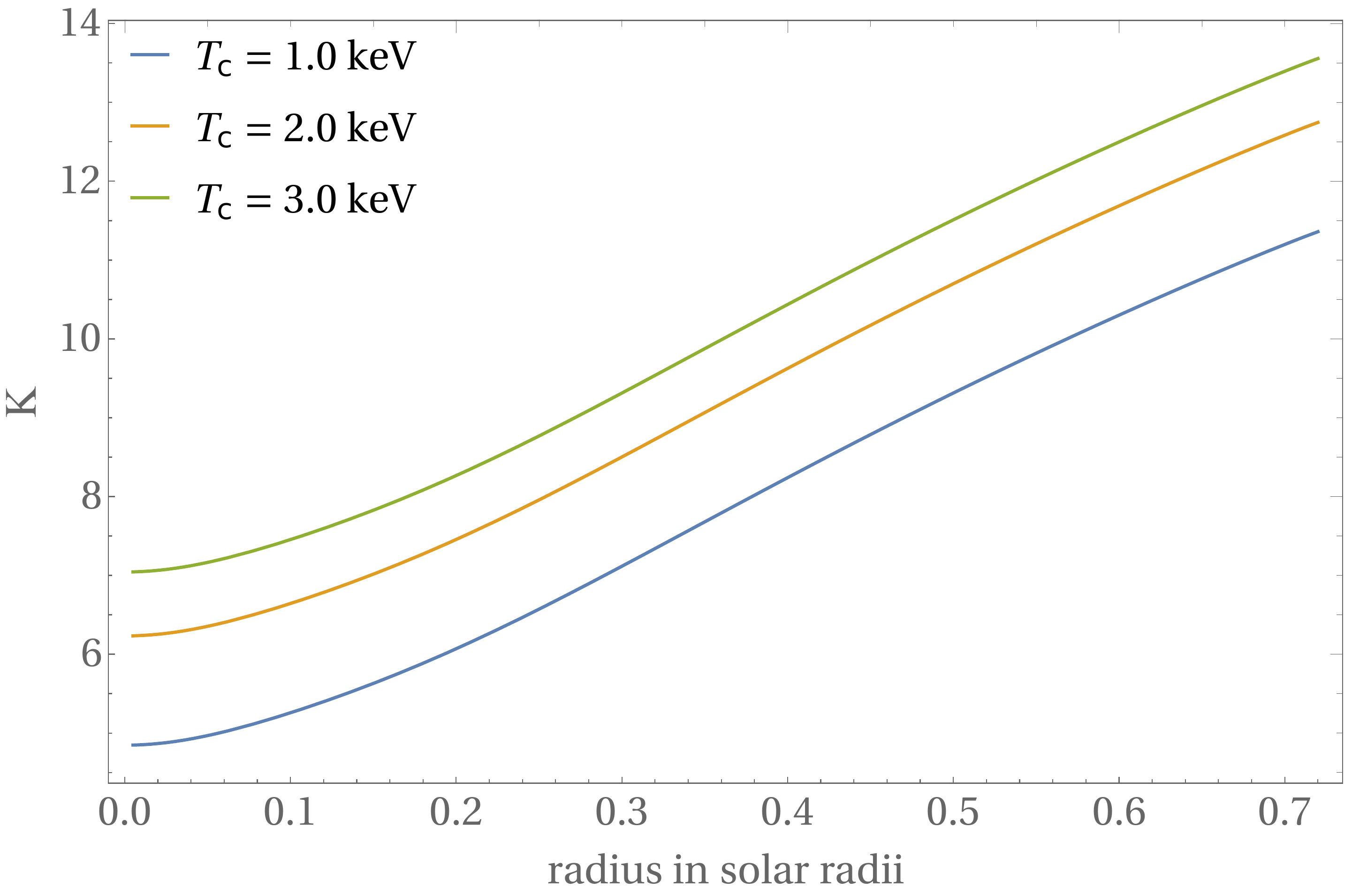}
\end{figure}

\begin{figure}[h!]
	\caption{Radial profile of the temperature of the millicharged gas $T_c$ inside the inner part of the radiative zone which provides the heating required to reconcile the B16 Standard solar model with helioseismology data; surface chemical composition AGSS09}\label{tcr}
	\centering \includegraphics[height=8cm]{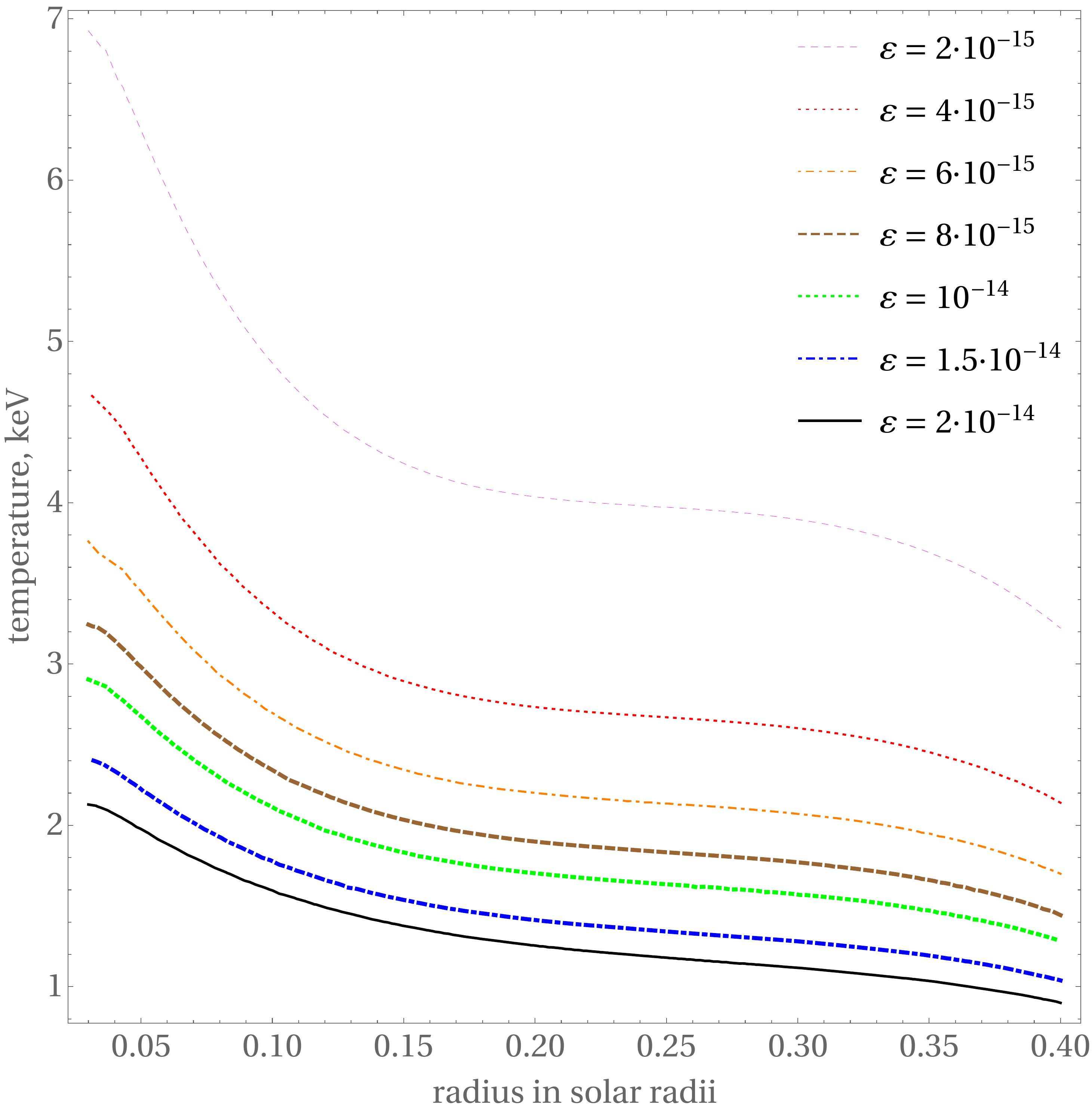}
\end{figure}

In order to calculate the energy transfer $Q$ we find an average with respect to the energies of the millicharged particles $\omega_c$:
\begin{equation}
\begin{split}
Q = 2\, n_e n_c \left\langle \int \Delta \omega_c \, d\sigma \right\rangle = \frac{\varepsilon^2 e^4 n_e n_c}{4 \, \pi m_e} \cdot \left\langle \left( \xi T - \omega_c \right) \cdot \frac{\ln \left( 4\, \omega_c^2 / \omega_{pl}^2 \right) - 1}{\omega_c} \right\rangle .
\end{split}
\end{equation}
Now we are ready to determine the constant $\xi$ taking advantage of the condition $Q = 0$ as soon as $T = T_c$. i.e. if electrons and millicharges are in thermal equilibrium. Finally we get:
\begin{equation}
Q =  \frac{\varepsilon^2 e^4 n_e n_c}{4\, \pi m_e} \cdot \left( \frac{T}{T_c} - 1 \right) \cdot \text{K} \left( \frac{4\, T_c^2}{\omega_{pl}^2} \right), \quad \text{K} \left( a \right) \equiv \frac{2}{3 \zeta\! \left( 3 \right)} \int\limits_0^{\infty} dx\, \frac{x^2 \left( \ln ax^2 - 1 \right)}{e^x+1}.
\end{equation}
The quantity $\text{K} \left( a \right)$ is showed in the Fig. \ref{K} as a function of radial coordinate for the three values of temperature of the millicharged particles.
Let us use $\omega_{pl}^2 = e^2 n_e / m_e$ and $n_c = 3 \zeta \! \left( 3 \right) T_c^3 / (2\,\pi^2)$ in order to write $Q$ in a more convenient form:
\begin{equation}\label{qq}
Q = 1.6\, \frac{\text{erg}}{\text{s} \cdot \text{cm}^3} \cdot \left( \frac{\varepsilon}{7\cdot 10^{-15}} \right)^2 \left( \frac{\omega_{pl}}{290\,\text{eV}} \right)^2 \left( \frac{T_c}{\text{keV}} \right)^3 \left[ \frac{T}{T_c} - 1 \right] \text{K} \left( \frac{4\, T_c^2}{\omega_{pl}^2}  \right) ,
\end{equation}
where $\omega_{pl} = 290\, \text{eV}$ is the plasma frequency in the centre of the Sun. If $T > T_c$, then the millicharged particles get energy from the plasma: $Q > 0$. Otherwise the energy is transferred from the millicharged particles to the solar plasma. Now we calculate the variation of the luminous flux power $\delta l_c$ corresponding to the energy transfer $Q$:
\begin{equation}
\delta l_c \! \left( r \right) = - \frac{4\pi R_{\odot}^3}{L_{\odot}} \cdot \int\limits_0^r Q\! \left( x \right) x^2 \, dx .
\end{equation}
In order to provide the required heating inside the solar core and resolve the solar abundance problem, we need this variation to match the one found in the section \ref{opt} and depicted in the Fig. \ref{dl1}. Thus we find the dependence of the temperature of the millicharged gas $T_c$ on the radial coordinate for any allowed value of the millicharge $\varepsilon$, see Fig. \ref{tcr}. The corresponding distribution of the magnetic field can be found given the equilibrium condition for the pressure: $B^2 / 2  = 7\pi^2\, T_c^4 / 180 \; \Rightarrow \; B = 45\,\text{MG} \cdot \left( T_c / \text{keV} \right)^2$. This is normally higher than the existing constraints on the magnetic field inside the solar core which we discussed in the section \ref{magnetic}, however, as we have already noted there, these constraints are really the constraints on the pressure excess, so that they are relaxed as long as one takes into account the pressure exhibited by the millicharged gas.

%------------------------------------------------

\section{Conclusion}

\begin{figure}[h!]
	\caption{Blue and yellow: change in the power of luminous flux required to solve the solar abundance problem; red rectangle: interval of radii where the new source of energy is needed; green rectangle (not to scale): radial coordinate where the new localized sink of energy is needed; black dashed line: change of luminous flux power in the physical model discussed (for the interval of radii where it does not coincide with the blue line)}\label{final}
	\centering \includegraphics[height=8cm]{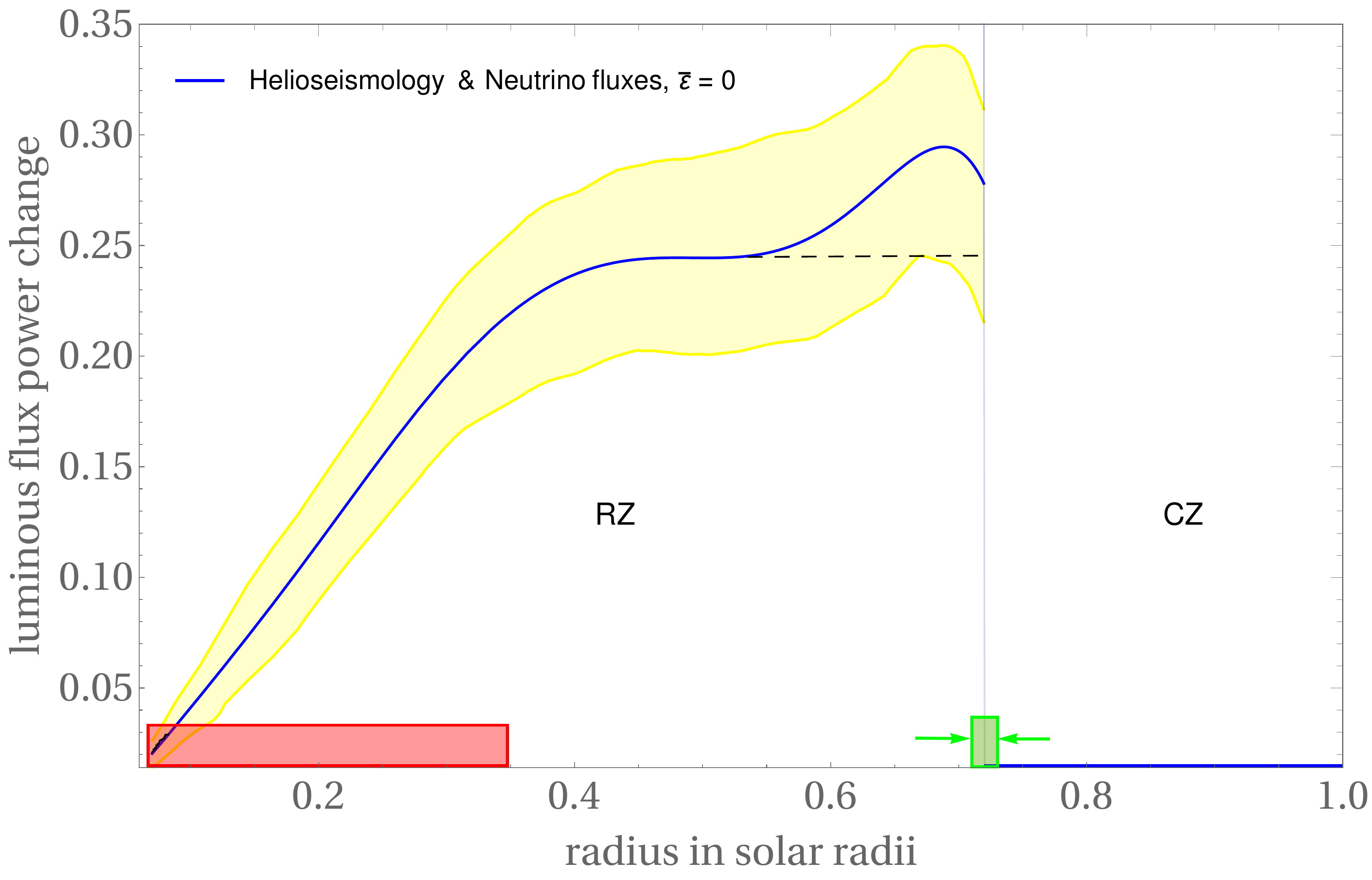}
\end{figure}

In this work we studied the solar abundance problem in the context of the hypothetical additional energy transfer inside the Sun. We analyzed the influence of the hypothetical additional energy sources or sinks inside the Sun on the solar abundance problem and figured out that the problem cannot be solved by neither emission nor absorption of energy alone. We showed that the solar abundance problem can be considered as an evidence of additional energy transfer inside the solar radiative interior. In order to find the change in energy transfer compared to the B16 Standard solar model, we calculated the radial profile of the luminous flux power which is evidenced by the combined data from the helioseismology, spectroscopy and solar neutrino detection experiments. The required addition to the luminous flux power was found (Fig.~\ref{dl1}), which can be used to build a physical model that solves the solar abundance problem. In particular, our investigations suggest that the anomaly is eliminated if there is a localized loss of energy near the radiative zone boundary and an absorption of approximately the same amount of energy inside the solar core ($0.1R_{\odot} \lesssim r \lesssim 0.3R_{\odot}$). We showed that if the localized loss of energy is associated with the resonant emission of transversely polarized hidden photons, the kinetic mixing parameter required does not contradict any of the existing constraints, the ones from the Sun \cite{Redondo:2013lna} being relaxed as long as the energy loss is compensated by energy deposition inside the core. To be more precise, the parameters of the model are:  kinetic mixing $\chi \simeq 10^{-13}$, hidden photon mass $m_{\gamma'} = 12\,\, \text{eV}$. Then we discussed the mechanism for the additional heating of the solar plasma of the inner part of the radiative zone. Within the scope of this paper, we decided to continue exploring the hidden photon model and study if it can provide the heating required. We supposed that there are light dark fermions charged under the hidden photon $U\left(1\right) '$ group. Given the energy scale characterizing their interactions in our case, these dark fermions acquire an effective millicharge $\varepsilon$. We showed that the magnetic fields of the solar radiative zone can capture millicharged particles with the energies $\omega_c \lesssim \left( 1 - 10 \right)\, \text{keV}$ and with the allowed values of the millicharge $\varepsilon \sim 10^{-15} - 10^{-14} $. Finally, we found the distribution of the millicharged particles inside the inner part of the radiative zone which is required to provide the heating of the solar plasma discussed earlier. We illustrate the regions of the overall contribution to the additional energy transfer from the hidden photon emission (green, not to scale) and the collisions of the millicharged particles with the solar plasma (red) in the Fig.~\ref{final}. The change of the luminous flux power in the model which we discuss coincides with the required change of the luminous flux power calculated in the section \ref{opt} of this work (blue line), except for the interval marked by a black dashed line. There are no sources or sinks of energy within the latter interval in our model, therefore the luminous flux power remains constant. One can see that the model we discuss explains the solar abundance anomaly quite well; however, there is a question of how the required distribution of the millicharged particles inside the solar core is maintained, which we address only briefly with the suggestion of a mechanism involving a source of dark photons in the solar centre. Another interesting possibility is that the required additional heating inside the core originates from a different kind of physics which is not related to the hidden photon models. For example, a change in S-factors of thermonuclear reactions can increase luminous flux inside the core. If such an increase is then compensated by the resonant hidden photon emission that we discussed in this work, the overall picture should look similarly to what is depicted in the Fig.~\ref{final}, thus possibly solving the solar abundance problem. 

The proposed additional energy transfer solution to the solar abundance problem has well determined implications for future experimental tests. Even a low accuracy measurement of the neutrino fluxes from the reactions of the solar CN-cycle can provide one with the abundances of volatiles inside the core, thus directly probing the chemical composition of the solar interior \cite{2013PhRvD..87d3001S}. If the observed metallicity of the interior is high, then the solar abundance problem has to be solved the other way, by explaining what is wrong with the current understanding of the solar chemical evolution or surface composition. Besides, it is possible that in the near future one finally detects g-modes of solar oscillations \cite{2010A&ARv..18..197A}, which will largely refine our knowledge about the inner structure of the Sun. Then one will see if the simple additional energy transfer outlined in this paper can still explain the solar abundance anomaly and if yes, what features this transfer must have. Finally, a very important role is to be played by future laboratory measurements of the opacity of different elements at solar temperatures, such as the recent experiments \cite{Bailey2016, PhysRevLett.122.235001}. These measurements, along with the further theoretical work on the refinement of opacity calculations, can largely decrease uncertainties in solar modeling and shed light on the details of the existing anomalies, providing an accurate profile of the additional energy transfer required to solve the solar abundance problem and enabling one to test the concrete physical models underlying this transfer.

%%%%%%%%%%%%%%%%%%%%%%%%%%%%%%%%%%%%%%%%%%%%%%%%%%%%%%%
\acknowledgments  I am indebted to  S.~Troitsky and D.~Gorbunov for numerous valuable discussions and comments on the manuscript. I thank DESY theory group for hospitality at the final stages of this work. The work was supported by the Russian Science Foundation grant 14-22-00161. 

\bibliography{thesis}

\end{document}